\documentclass[aps,prf,twocolumn,showpacs,longbibliography]{revtex4-1}
\usepackage{amsfonts}
\usepackage{amsmath}
\usepackage{amsthm}
\usepackage{graphics}
\usepackage{graphicx}
\usepackage{subfigure}
\usepackage{amssymb}
\usepackage{graphics}
\usepackage{graphicx}
\usepackage{epstopdf}
\usepackage{color}
\usepackage{subfigure}
\usepackage{pgfplots,tikz}
\usepackage{url}
\usepackage{soul}
\usepackage{enumitem}
\usepackage{lipsum}
\usepackage{hyperref}
\usepackage{cleveref}

\newcommand\bet{{g}}
 
\newcommand\alps{{\frac{\hbar^2}{2m}}}

\newcommand\dertt[1]{ \frac{\partial{ #1}}{\partial t} }
\newcommand\gd{\mbox{${\bf \nabla}^{2}$}}

\begin{document}

\title{(Non)-universality of vortex reconnections in superfluids}

\author{Alberto Villois}
\affiliation{School of Mathematics, University of East Anglia, Norwich Research Park, Norwich, NR4 7TJ, United Kingdom}
\author{Giorgio Krstulovic}
\affiliation{Universit\'e de la C\^ote d'Azur, OCA, CNRS, Lagrange, France. B.P. 4229, 06304 Nice Cedex 4, France}
\author{Davide Proment}
\affiliation{School of Mathematics, University of East Anglia, Norwich Research Park, Norwich, NR4 7TJ, United Kingdom}


\begin{abstract}
An insight into vortex reconnections in superfluids is presented making use of analytical results and numerical simulations of the Gross--Pitaevskii model.
Universal aspects of the reconnection process are investigated by considering different initial vortex configurations 
and making use of a recently developed tracking algorithm to reconstruct the vortex filaments.
We show that during a reconnection event the vortex lines approach and separate always accordingly to the time scaling $ \delta \sim t^{1/2} $ with pre-factors that depend on the vortex configuration. 
We also investigate the behavior of curvature and torsion close to the reconnection point, demonstrating analytically that the curvature can exhibit a self-similar behavior that might be broken by the development of shock-like structures in the torsion. 
\end{abstract}
\maketitle
\section*{Introduction}
Reconnections in fluids have been object of study for long time in the contest of plasma physics \cite{Priest1999} and both classical \cite{Kida1994}
and superfluid dynamics \cite{KoplikLevinPRL1993}. 
Depending on the physical system considered, such reconnections are events characterized by a rearrangement in the topology of either magnetic field, (magnetic reconnections) or vorticity field (vortex reconnections).
Such topological modifications are believed to play a fundamental role in several physical phenomena like eruptive solar events \cite{Zhike2016}, energy transfer and fine-scale mixing \cite{Fazle&KarthikPof2011} and turbulent states in superfluids \cite{Fonda25032014}. 
Despite their physical relevance, reconnections represent also a stand-alone mathematical problem, related for instance, to the presence of singularities in the Euler equation \cite{Kida1994, constantin1996geometric, Moffatt2000}.

In classical fluids described by the Navier--Stokes-type equations, reconnecting vortex tubes stretch and deform, leading to complicated dynamics and formation of structure like vortex bridges \cite{Fazle&KarthikPof2011}.
In order to understand fundamental aspects of vortex reconnections it is often desirable to work with a vortex configuration where the vorticity results confined along lines of zero core size. 
Such idealization is called a vortex filament.
This limit naturally arises in superfluids, such as superfluid liquid Helium (He II) and Bose--Einstein condensates (BECs). 
Superfluids are in fact examples of ideal flows of quantum mechanical nature characterized by the lack of viscous dissipation and by a Dirac's $ \delta $ vorticity distribution supported on the vortex filaments.
For such fluids, the velocity circulation is equal to a multiple of the Feynman-Onsager quantum of circulation $ \Gamma=h/m$, with $h$ the Planck constant and $m$ the mass of the superfluid's bosonic constituents.

Due to the Kelvin's circulation theorem (or the Alfv\`en's theorem in magneto-hydrodynamics), in a barotropic ideal flow reconnections should be forbidden since the circulation of vortex lines transported by the flow is conserved and so their topology is frozen.
However, as already suggested by pioneering works of Feynman \cite{Feynman195517} and Schwarz \cite{Schwarz_VF}, vortex reconnections in superfluids do exist and play a fundamental role in superfluid turbulence.
This was indeed confirmed in the early 90's by Koplik and Levine \cite{KoplikLevinPRL1993} who performed numerical simulations of reconnecting vortex lines within the Gross--Pitaevskii (GP) model.
They showed that the Kelvin's circulation theorem does not hold in this context because the superfluid density identically vanishes at the vortex filament.
With the progress of experimental techniques in the last decade, reconnecting superfluid vortices have been visualized in He II \cite{BewleyReconnectionPNAS, Fonda25032014} and in BECs \cite{PhysRevLett.86.2926, PhysRevLett.115.170402}.
From the theoretical side, many works have been devoted to study the reconnecting vortex filaments in superfluids, either by using the so-called vortex filament (VF) model introduced by Schwarz \cite{Schwarz_VF} or the GP model.

The simplest question to ask, although contradictory answers appear in the literature, is related to the rates of approach and separation of two reconnecting vortices. 
Assuming that a reconnection event is a local process in space and the circulation $ \Gamma $ is the only relevant dimensional quantity involved,
by simple dimensional analysis it follows that the distance $\delta(t)$ between two reconnecting filaments should scale as
\begin{equation}
\delta(t)\sim (\Gamma \, t)^{1/2} \, ,
\label{EQ:scalingDelta}
\end{equation}
independently if it is measured before or after the reconnection.
Such prediction has been confirmed by numerical simulations of the VF model \cite{Baggaley&Sherwin&Barenghi&SergeePRB2012} and in He II experiments \cite{BewleyReconnectionPNAS}.
In the framework of the GP model, the same scaling was asymptotically derived in \cite{nazarenko2003analytical} but a number of numerical studies report disparate scaling exponents that may differ between the before and after reconnection stages \cite{Zuccher&Caliari&Baggaley&BarenghiPof2012,AllenFiniteTempRecoPRA2014,2014arXiv1410.1259R}.
Another fundamental question regards the universality of the geometrical shape of the vortex filaments at the reconnection.
It is expected that vortices become locally antiparallel during the reconnection process \cite{nazarenko2003analytical}. 
However, using the VF model it has been reported that the reconnection angle may follow a broad distribution that depends on turbulent regime that is considered \cite{Baggaley&Sherwin&Barenghi&SergeePRB2012}.
It has been also observed that during a reconnection event cusps are generated on the filaments and argument has been given either in favor of those cusps being universal \cite{Waele&AartsPRL1994} or not \cite{tebbs2011approach}.
Finally, a lot interest has arisen recently on the generation of Kelvin waves (helical waves propagating along vortex filaments) \cite{PhysRevLett.86.3080, Nazarenko2007, Fonda25032014, PhysRevB.92.184508} and evolution of hydrodynamical helicity \cite{Scheeler28102014PNASDavide, kimura2014, laing2015conservation, Zuccher&RiccaPRE2015, diLeoniHelicity} during reconnection events.

The VF model is based on Biot-Savart equations that describe a regularized Dirac's $ \delta $ vorticity distribution field in the incompressible Euler equation; it provides direct information on the vortex filaments and it is widely used to mimic superfluid vortex dynamics and turbulence in He II.
However, due to Kelvin's circulation theorem in the Euler equation, reconnections need here to be added by some ad-hoc cut-and-connect mechanisms.
In addition the VF model introduces a small scale cut-off to regularize Biot-Savart integral divergence and thus cannot explore the vortex dynamics at the smallest scales where the reconnection events take place.
The GP model represents an alternative in the study vortex dynamics and reconnections, the main advantages being that it naturally contains vortex reconnections in its dynamics and that the entire reconnection process is regular due to the identically zero superfluid density field at the vortex core.
Studying such small scale dynamics is crucial for understanding how energy is transferred through scales and eventually dissipated. 
Unfortunately, no information can be directly inferred from GP on the vortex dynamics because this model described the evolution of an {\it order parameter} complex field which contains simultaneously sound excitations and vortex lines in the form of topological defects. 
We will present here a detailed study of vortex reconnections by exploiting a recently developed tracking algorithm \cite{VilloisTrackingAlgo} that is able to track vortex filaments in numerical simulations of the GP model with an machine epsilon level of accuracy.

In order to understand what is universal in vortex filament reconnection mechanisms, we study the dynamics of four different initial configurations: (a) perpendicular and (b) almost anti-parallel lines, (c) a trefoil knot, and (d) reconnections occurring in a full turbulent tangle dynamics.
We will show that reconnecting vortex lines always obey the dimensional analysis scaling \eqref{EQ:scalingDelta} (both before and after reconnection), and they generally separate faster than they approach. 
In addition we report that regardless of the initial configuration, vortices become anti-parallel at the reconnection. 
We also report a self-similar behavior of the curvature close to the reconnection point when torsion does not play an important role and shock-like structures appearing in the torsion evolution for some configurations.
Those findings are explained by some asymptotic calculations.

\section*{The GP model and the reconnection case studies}
The GP model is a dispersive non-linear wave equation describing the dynamics of the order parameter $ \psi $ of a BEC arising in dilute Bose gases; for the sake of completeness and clarity we introduce it in App.~\ref{SubSec:GP}.
When $\psi$ is linearized about a constant value $\psi_0= \sqrt{\rho_0/m} $, the sound velocity results in $ c=\sqrt{g \rho_0}/m $ and dispersive effects take place at length scales smaller than the healing length $ \xi= \hbar / \sqrt{2 \rho_0 g} $.  
This can be easily understood by rewriting the GP model using those physical parameters
\begin{equation}
i\dertt{\psi} = \frac{c}{\sqrt{2}\xi}\left(-\xi^2\nabla^2\psi+\frac{m}{\rho_0}|\psi|^2\psi \right) \, 
\label{Eq:GPhydro}
\end{equation}
and comparing the magnitude of the r.h.s. first and second terms.  
Note also that by a suitable time and space rescaling the parameters $c$ and $\xi$ can be re-absorbed; in this work length- and time-scales are expressed in units of the healing length $\xi$ and its characteristic time $\tau=\xi/c$. 

The relationship between the GP equation and a hydrodynamical model is immediately illustrated by introducing the Madelung's transformation
\begin{equation}\label{eq:Madelung}
\psi({\bf x},t)=\sqrt{\frac{\rho({\bf x},t)}{m}}e^{i \frac{\varphi({\bf x},t)}{\sqrt{2}c\xi} },
\end{equation}
that relates $\psi$ to an inviscid, compressible, irrotational and barotropic superfluid of density $\rho({\bf x},t)$ and velocity ${\bf v}={\bf \nabla} \varphi$.
In the domain where the Madelung's transformation is well defined ($\psi\neq0$), the velocity field is potential. However, vortices may exist as topological defects of the order parameter. 
In places where the density vanishes (nodal lines) $ \arg \psi $ is not defined. 
The field $ \psi $ still remains a single-valued function if the circulation $ \oint \bf v \cdot d\bf \ell $ along a nodal line is a multiple of the quantum of circulation $\Gamma=h/m=2\sqrt{2}\pi c\,\xi$.
For this reason nodal lines of $ \psi $ are called quantum (or quantized) vortices. Their corresponding velocity field $\bf v$ thus decay as the inverse of the distance to the vortex, and their vorticity is therefore a Dirac-supported distribution.
Their typical vortex core size is order of $ \xi $.

\begin{figure*}
\centering
\includegraphics[width=1.025\linewidth]{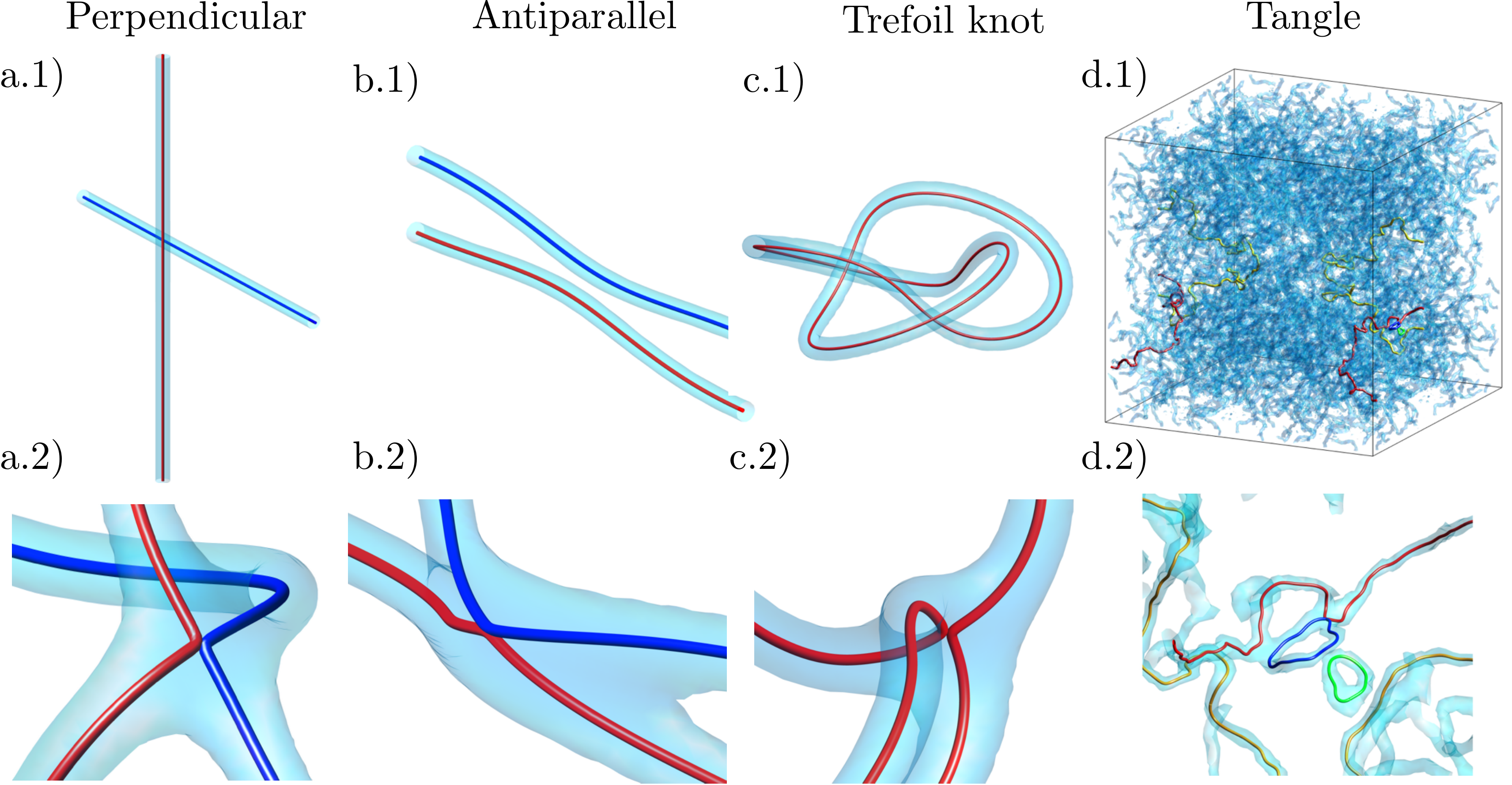}
\caption{(Color online) 3D plot showing the reconnection events explored numerically. The initial configuration is displayed for the perpendicular vortex lines (a.1), the anti-parallel lines (b.1) and the trefoil know (c.1). Figures (a.2), (b.2), (c.2) show a corresponding zoom at the moment of reconnection. Figure (d.1) displays the turbulent tangle and (d.2) a zoom in a place where a reconnection takes place. Red and blue correspond to the reconnecting vortex filaments, the light blue iso-surfaces render the density field at low values.}
\label{Fig:Lines}
\end{figure*}

The GP equation \eqref{Eq:GPhydro} is numerically integrated with a pseudo-spectral code. The resolution is chosen carefully to sufficiently resolve the vortex core in space and the reconnections in time.
We consider four different initial configurations in a cubic box of size $L$ with $N$ collocation points in each dimension:

\begin{description}[leftmargin=.0cm]
\item[a) Perpendicular lines.] 
The order parameter field characterized by straight vortex filaments perpendicular to each other and having initial distance of $ 6\xi $.
This initial configuration is shown in Fig.\ref{Fig:Lines}a.1.
$ L/\xi=128, N=256 $.
\item[b) Antiparallel lines.] Vortex filaments with opposite circulation are set at an average distance of $ 6\xi $. In order to trigger a Crow instability \cite{Berloff2001} a small perturbation is introduced by adding a Kelvin wave of amplitude $ \xi $ and wavelength equal to the system size. 
The initial configuration is shown in Fig.\ref{Fig:Lines}b.1.
$ L/\xi=128, N=256 $.
\item[c) Trefoil knot.] 
A vortex filament reproducing a torus $ \mathcal{T}_{2, 3} $ knot (a trefoil) is produced following \cite{Proment2012}; the torus on which the knot is built has toroidal and poloidal radii of $ R_0 = 16\xi $ and $ R_1 = 4\xi $ respectively.
The initial configuration is shown in Fig.\ref{Fig:Lines}c.1.
$ L/\xi=128, N=256 $.
\item[d) Turbulent tangle.]  
We prepare an initial condition consisting of several large-scale vortex rings that replicates a Taylor--Green flow as in \cite{nore1997decaying}.
The initial condition is then evolved in time: the rings reconnect breaking the initial symmetry and creating a dense turbulent tangle displayed in \ref{Fig:Lines}d.1 (see \cite{VilloisTangleLetter} for complete description of the field evolution).
We study four successive vortex reconnection events occurring in a small volume (Fig.\ref{Fig:Lines}d.2) at stages when the tangle density is higher.
$ L/\xi=256, N=256 $.
\end{description}
The time stepping scheme for cases (a) and (b) is a Strang-spliting method, whereas for cases (c) and (d) is a second-order Runge-Kutta. In each case, the time step is chosen to be smaller than the fastest linear time scale of the system. Conservation of the invariants has been carefully checked.

\section*{Approach and separation rates}
\begin{figure*}
\centering
\includegraphics[width=0.245\textwidth]{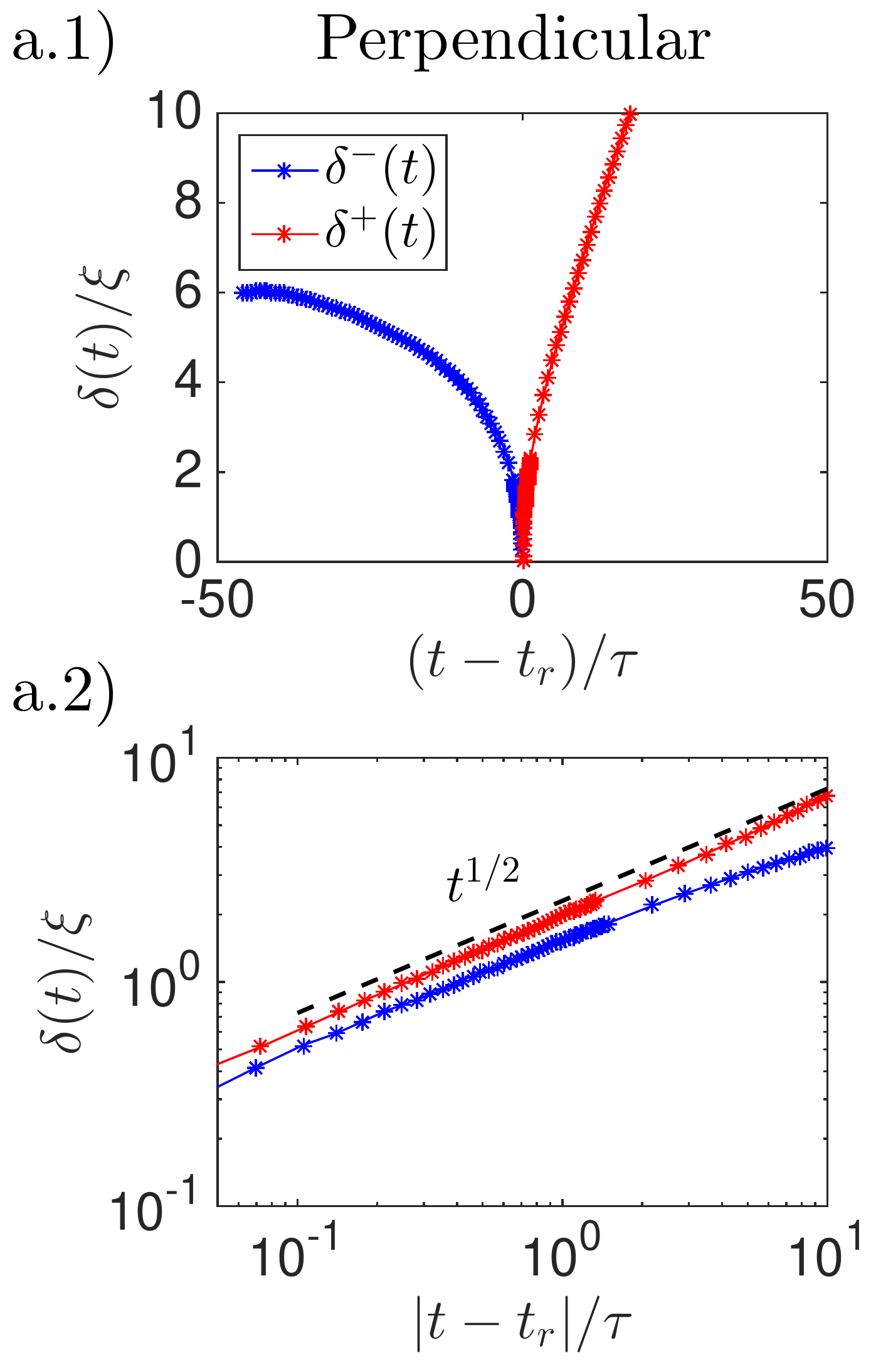}
\includegraphics[width=0.245\textwidth]{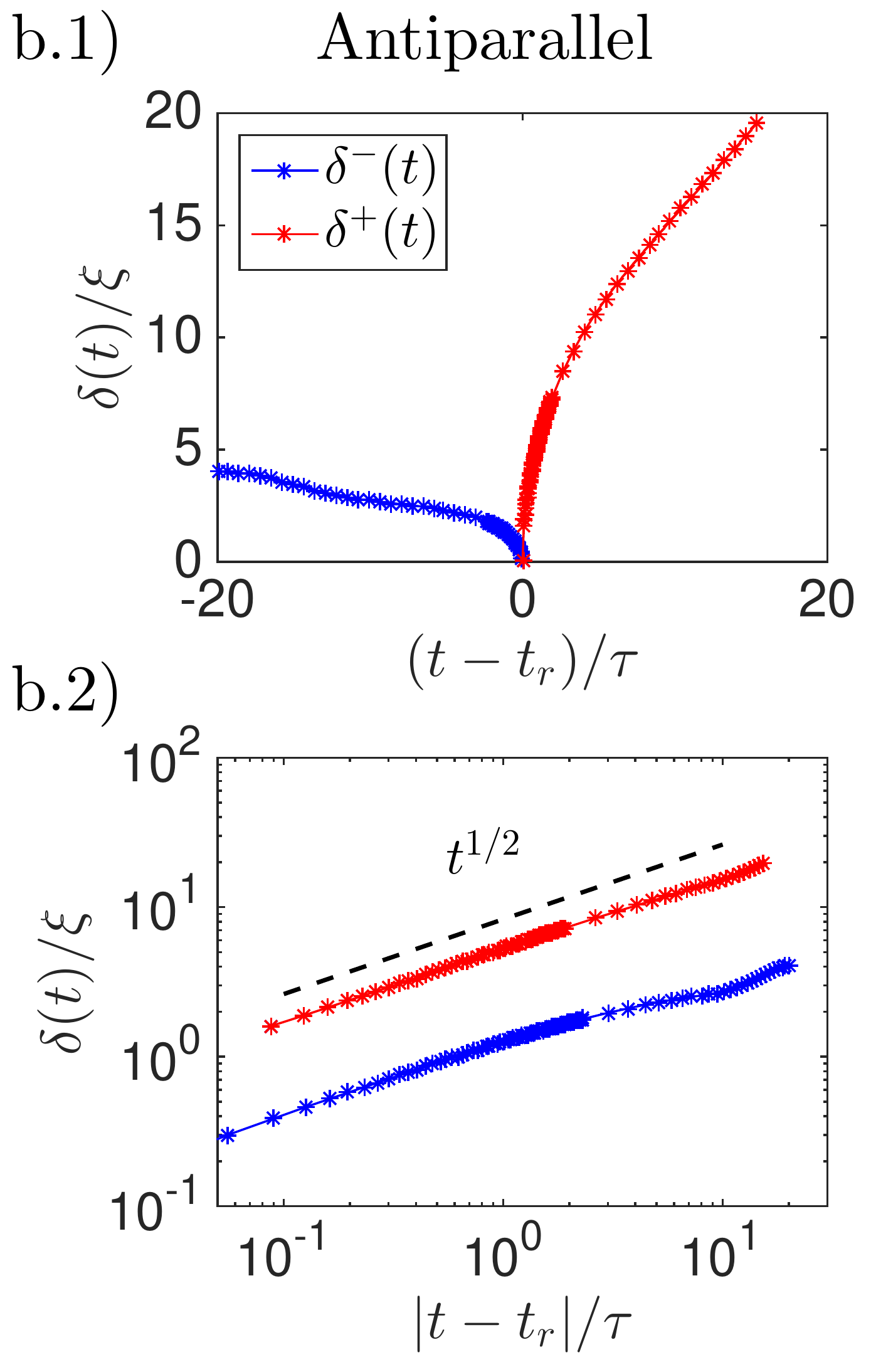}
\includegraphics[width=0.245\textwidth]{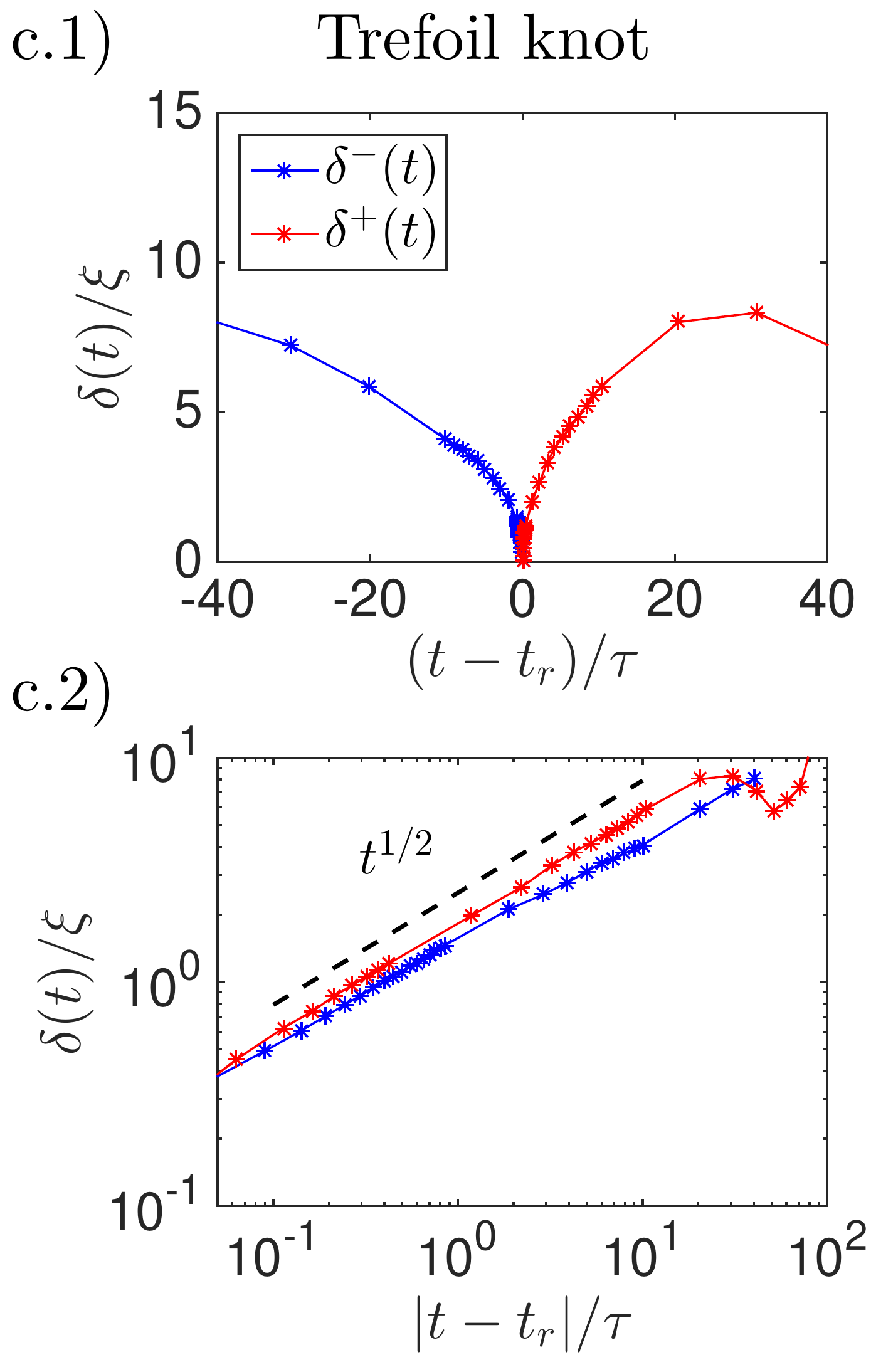}
\includegraphics[width=0.245\textwidth]{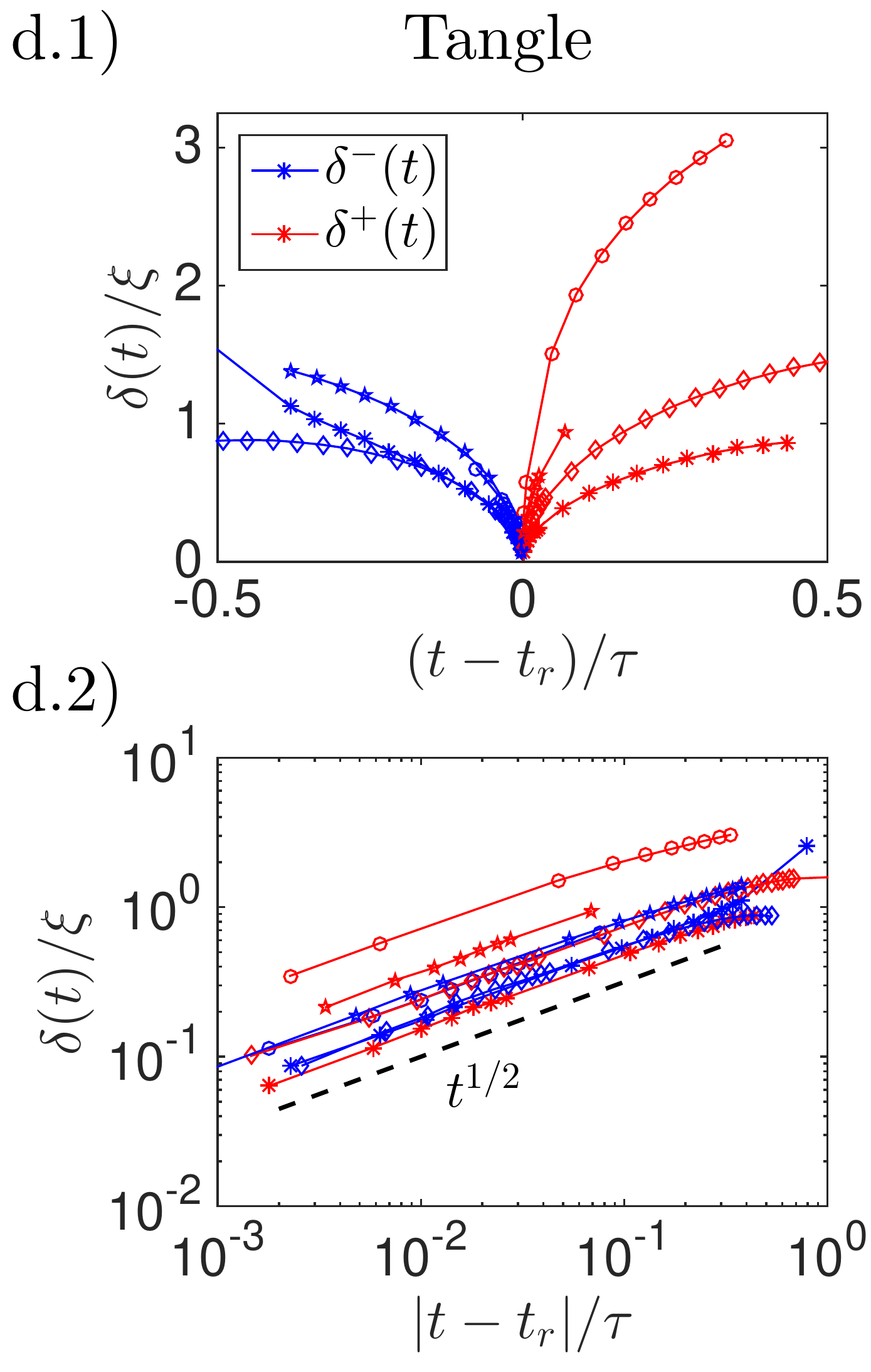}
\caption{(Color online) Temporal evolution of the distance between the reconnecting vortex filaments before (blue) and after (red) the estimated reconnection time $t_{r}$ for the perpendicular (a), antiparallel (b), trefoil knot (c) and turbulent tangle (d) configurations. For the turbulent tangle four different reconnection events have been tracked. a.2-d.2) same plots as in a.1-d.1 but in LogLog scales. 
}
\label{Fig:Delta}
\end{figure*}

Apart from the characteristic length scale $ \xi $ inherently present in the GP model, when quantized vortices are considered, the quantum of circulation $ \Gamma $ can be used to formulate an extra length scale. 
Hence, by dimensional analysis the distance between two reconnecting lines is expected to be
\begin{equation}
\delta^\pm(t)=A^\pm \, \xi^{1-2\alpha^\pm}| \Gamma \, (t-t_r)|^{\alpha^\pm},
\label{eq:delta}
\end{equation}
where $ \alpha^\pm $ and $A^\pm$ are dimensionless parameters and the superscript $\pm$ stands for before (-) and after (+) the reconnection event. 
The temporal evolution of the minimal distances between reconnecting filaments for the different case studies are displayed in Fig.\ref{Fig:Delta}a-d.
An explanatory movie of the knot reconnection is also provided as SI.
Remarkably, in all cases the approach and separation rates follow the same dimensional $ t^{1/2} $-scaling. 
For each event we estimate the reconnection time $t_r$ by doing a linear fit on $\delta^{\pm}(t)^2$ and compute $ t_r $ as the arithmetic mean between $ t_r^{\pm} $ that satisfy $ \delta^{\pm}(t_r^{\pm})^2=0 $. 
%Let us remark that the tracking algorithm allows us to extend the distance measure to length-scales smaller than the vortex core size and the computational grid. 
The $ t^{1/2} $-scaling extends beyond $ \xi $ and only slight deviations are observed in some cases.
Perhaps this fact could explain the different results for the scaling obtained in \cite{Zuccher&Caliari&Baggaley&BarenghiPof2012,AllenFiniteTempRecoPRA2014,2014arXiv1410.1259R} where it was concluded that the exponents before and after the reconnection are different. 
For instance in \cite{Zuccher&Caliari&Baggaley&BarenghiPof2012} it was found that $\alpha^-\in(0.3,0.44)$ and $\alpha^+\in(0.6,0.73)$ and in \cite{2014arXiv1410.1259R} that either $\alpha^\pm=1/2$ or $\alpha^-=1/3$ and $\alpha^+=2/3$ depending on the initial vortex filament configuration.
In these works the time asymmetry was interpreted as a manifestation of the irreversible dynamics due to sound emission; we will come back to this interesting point in the Discussion.
Let us stress that the tracking algorithm we used is able to measure the inter-vortex distances even in presence of sound waves (the Taylor-Green tangle analyzed contains moderate sound at all scales) and no asymmetry concerning the exponent is observed.

\begin{figure}
\centering
\includegraphics[width=0.95\linewidth]{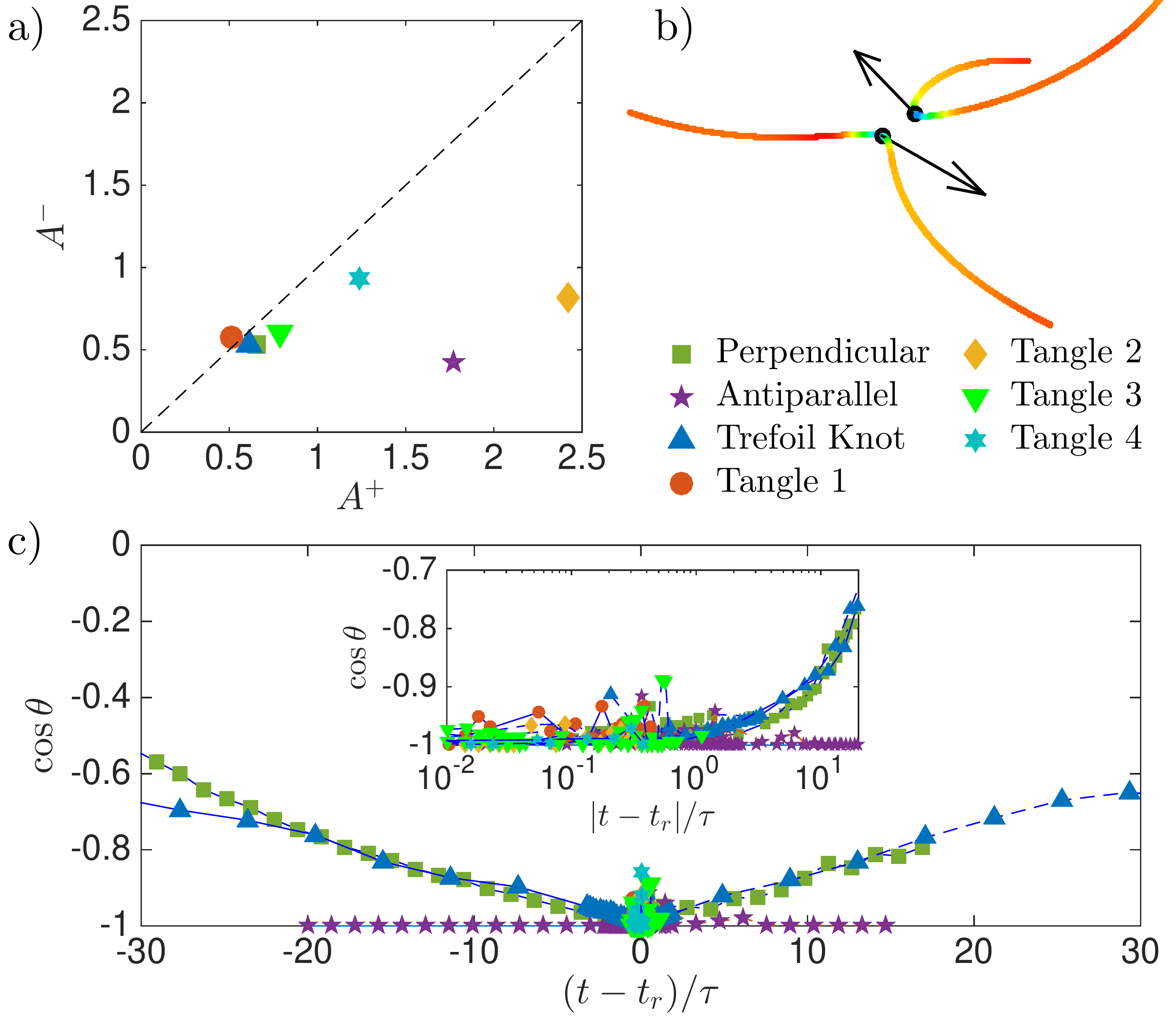}
\caption{(Color online) a) Fitted values of the pre-factors $A^\pm$ corresponding to \eqref{eq:delta}. b) An example of reconnecting filaments (trefoil knot case): the black dots represent the points of minimal distances and are used to compute $\delta(t)$, the arrows are the tangents of the filaments at those points, the reconnection angle $ \theta $ is defined by using the scalar product of the tangents. The coloring is proportional to the filament curvature (low in red and high in green/blue). c) Temporal evolution of the cosine of the reconnecting angle. The inset displays the same plot in LogLog scales.
}\label{Fig:Unic}
\end{figure}

Although the measured exponent is always $ \alpha^{\pm}=1/2 $, the full dynamics is not symmetrical with respect to the reconnection time as it can be immediately deduced by observing Fig.\ref{Fig:Delta}.
By estimating the pre-factors $ A^{\pm} $ with a fit, shown in Fig.\ref{Fig:Unic}a, we conclude that these are always order of the unity but are not universal.
Moreover, we observe that the vortex filaments usually separate faster than they approach ($A^- \lesssim A^+ $).

The tracking algorithm we use follows the pseudo-vorticity and it naturally provides the orientation of the filament with respect to the circulation. 
It thus allows us to compute the tangent vectors to the lines and infer the orientation of the filaments by evaluating the cosine of the angle $ \theta $ between the vectors at the two closest points as illustrated in Fig.\ref{Fig:Unic}b and in the supplied movie.
By approaching to the reconnection point each vortex filament develops a cusp-like structure characterized by high and localized values of the curvature (displayed in green/blue colors). 
The temporal evolution of $ \cos \theta $ for all the case studies is presented in Fig.\ref{Fig:Unic}.c. 
It is apparent that, independently of the initial configurations, vortices are always antiparallel at the reconnection point.
This behavior appears to be time-symmetric about the reconnection time and is smooth, as highlighted in the inset of Fig.\ref{Fig:Unic}.c where we show $ \cos \theta $ in LogLin coordinates for a better view on the short times before and after reconnection. 
%This is a first evidence that cusps-like structures in the filaments appear before and after the reconnection process as we shall demonstrate in the following.

\section*{Analytical predictions using a linear approximation}

The results presented in Figs. \ref{Fig:Delta} and \ref{Fig:Unic} support the analytical predictions obtained by Nazarenko\&West in \cite{nazarenko2003analytical}. 
Their seminal calculations consider a planar reconnection of two vortex filaments having a hyperbolic configuration at times close to $t_r$. 
As we shall observe in the following, vortex reconnections not always fully lay in a plane and the local torsion of the filament can play an important role. 
We generalize here the calculations performed in \cite{nazarenko2003analytical} including torsion of the vortex lines to understand its effect during the reconnection. 
Let us assume that at the reconnection time $t_r$ and close to the reconnection point the order parameter of two reconnecting non-planar vortex lines is given by
\begin{equation}
\psi_r(x,y,z)=z+\frac{\gamma }{a}(x^2+y^2)+i (a z+\beta  x^2- y^2),
\end{equation}
with $a\neq 0$ and $\frac{\beta-\gamma}{\gamma+1}>0$ (Nazarenko\&West reconnecting vortex profile is recovered by setting $\gamma=0$). 
In the vicinity of the vortex filaments, $ \psi $ is small and the non-linear term in \eqref{Eq:GPhydro} can be neglected. 
Within this approximation the pre- and post-reconnection solution is given by $\psi(x,y,z,t)=e^{ \frac{i (t-t_r)\Gamma}{4\pi}\nabla^2}\psi_r(x,y,z)$. 
By solving $\psi(x,y,z,t)=0$ we can explicitly obtain the temporal evolution of the vortex lines.
\eqref{eq:delta} is obtained with
\begin{equation}
\alpha^{+}=\alpha^{-}=\frac{1}{2} \quad \mbox{and} \quad
\frac{A^+}{A^-}=\sqrt{\frac{1+\gamma}{\beta-\gamma}} \, ,
\end{equation}
for $a>0$ and $ \beta<1-2\gamma/a^2$ (please refer to App.~\ref{SubSec:LinearApprox} for a figure of the vortex profiles, details on the above calculations and different choices of $ a $ and $ \beta $). 
Interestingly, the angle between the asymptotes of the hyperbolic vortex configuration close to reconnection is found to be $\phi=2\tan^{-1}(A^-/A^+)$.

The linear approximation also allows for computing the curvature and torsion of the vortex lines. As pointed out by Schwarz in \cite{Schwarz_VF}, the curvature $ \kappa^\pm(s, t) $ should present a self-similar behavior close to the reconnection point of the form $ \kappa^\pm(s, t) =\kappa^\pm_{\rm max}(t)\Phi^\pm(\zeta^\pm)$, where $\zeta^\pm=(s-s_r)\kappa^\pm_{\rm max}(t)$, $s_r$ is the coordinate of the reconnecting point and $ \kappa_{\rm max} $ is the maximum value of curvature. 
The present calculations predict 
\begin{equation}
 \kappa^\pm_{\max}(t) \propto |t-t_{ r} |^{-1/2} \,\, {\rm and} \,\, \frac{\kappa^+_{\max}(t)}{ \kappa^-_{\max}(t)}=\left(\frac{A^+}{A^-}\right)^{3}.\label{Eq:kappaMax}
\end{equation}
Note that the $ t^{-1/2} $-scaling could be directly inferred by dimensional analysis arguments but not the scaling of the dimensionless pre-factors.
Moreover these self-similar functions $\Phi^\pm(\zeta^\pm)$ can be expressed in compact forms for small values of $\gamma$ and $ (t-t_r) $ as:
\begin{equation}
\Phi^\pm(\zeta)=\frac{1}{\left\{1+\left[\left(\frac{{A^\mp}}{{A^\pm}}\right)^2 +1\right] \zeta^2\right\}^{3/2}} +O\left[\eta^\pm\gamma^2\frac{(t-t_r)}{\tau} \right],
\label{eq:curvSelfSim}
\end{equation}
with $\eta^\pm=  \left({A^\mp}/{A^\pm}\right)^2 -1$.
This function corresponds to a cusp in the vortex filament at $ t=t_r $ and $ s=s_r $.
The dependence on the coefficient $ \left({A^\mp}/{A^\pm}\right)^2 +1 $ multiplying the self-similar variable $ \zeta^\pm $ is unexpected and could not also be guessed by dimensional arguments.  
We also remark that the self-similarity is only exact when $ \gamma=0 $ or $ \eta^\pm = 0 $.

Finally, the torsion $\mathcal{T}^\pm(s,t)$ of vortex line can be also computed within this approximation.
When $ \gamma \neq 0 $ torsion is not identically null but it vanishes at $ s_r $, thus confirming that reconnections occur locally on a plane.
Also, it can be proved that it changes sign linearly at $ s_r $ with a slope that diverges as $\gamma|t-t_r|^{-1/2}$, creating shock-like structures.
The slope ratio before and after the reconnection satisfies the relation $\left.\frac{\mathrm{d}\mathcal{T}^+}{\mathrm{d}s}/\frac{\mathrm{d}\mathcal{T}^-}{\mathrm{d}s}\right|_{s=s_r}=A^+/A^-$. 

We observe that in the context of Euler and Navier--Stokes flows, dynamical equations for torsion and curvature have been derived in \cite{PhysRevE.51.3207}. 
To our knowledge, these non-linear equations do not allow for predicting the generation of curvature cusps and shock-like torsion structures. 
It would be interesting to investigate if the scaling laws reported above also remain valid in classical fluids and MHD flows.

\section*{Numerical measurements of the curvature and torsion}

\begin{figure}
\centering
\includegraphics[width=0.95\linewidth]{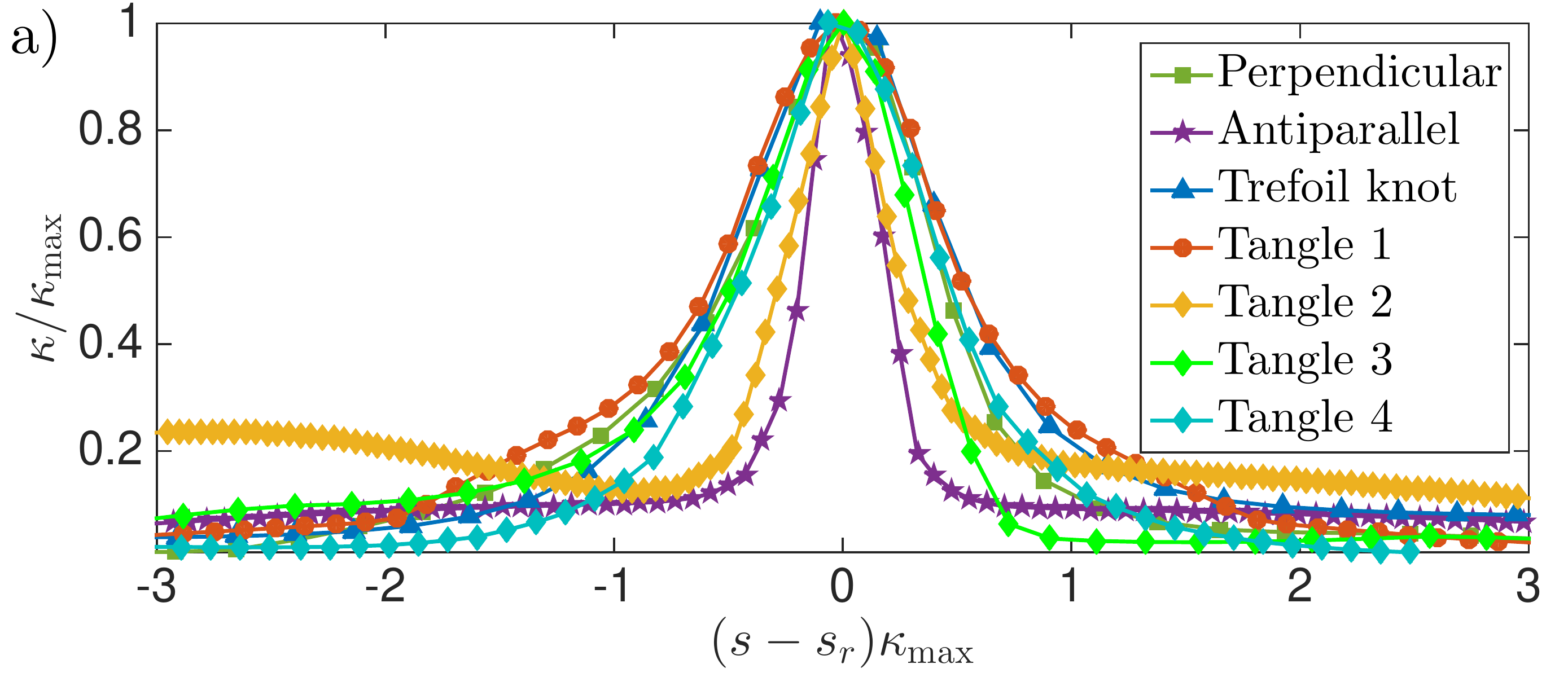} 
\includegraphics[width=0.95\linewidth]{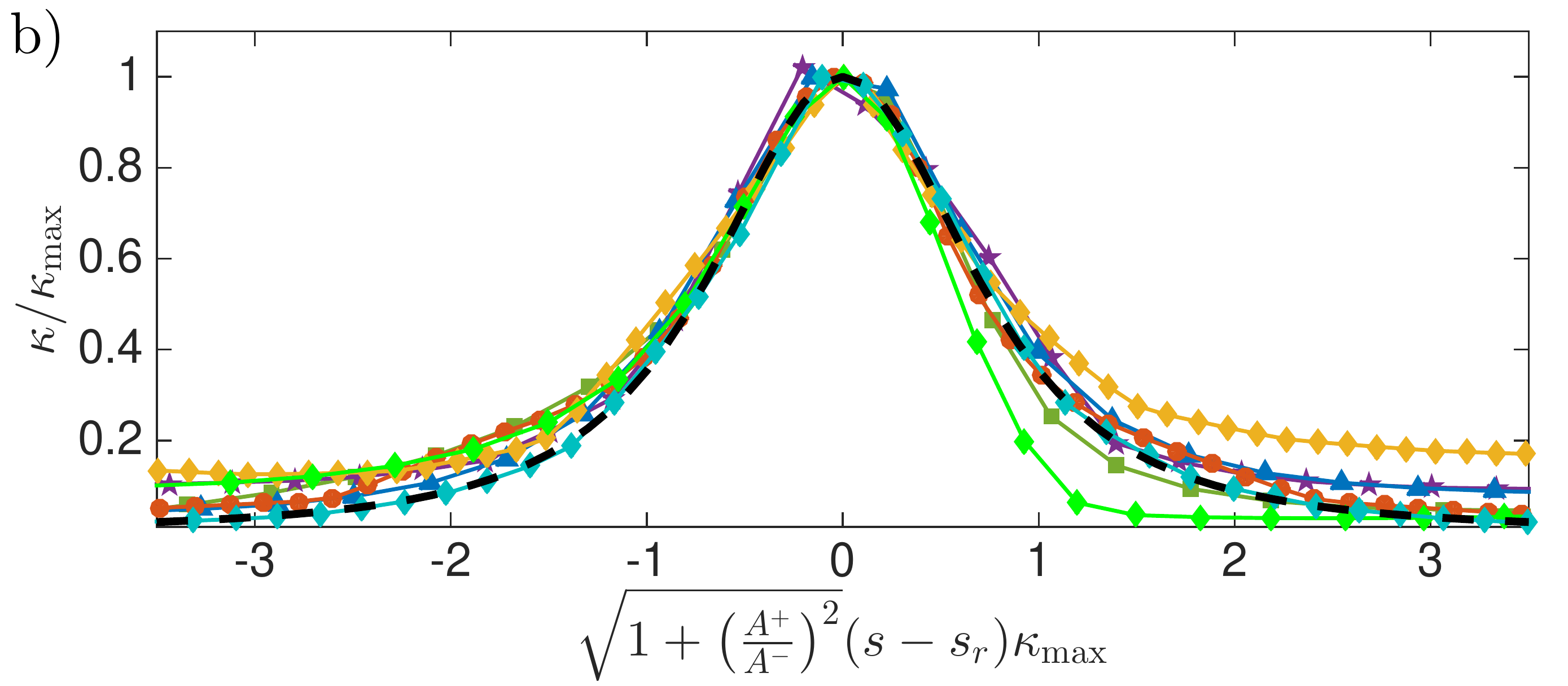} 
\caption{(Color online)  
 a) Curvature normalized by $\kappa_{\rm max}$ close to the reconnection time (just before) for all reconnection events explored as a function of $ (s-s_r) \kappa_{\rm max} $. b) Same data represented using the scaling suggested by the self-similar form (\ref{eq:curvSelfSim}). The black dashed line displays the theoretical prediction.}
\label{Fig:Cusps}
\end{figure}

Motivated by the previous asymptotic results we analyze the data coming from simulations. 
%Our tracking algorithm directly allows us for such numerical computations.
We start looking at the curvature at a fixed time very close to the reconnection. 
In Fig.\ref{Fig:Cusps}.a the curvature just before $t_r$ normalized using $\kappa_{\max}$ is shown for all configurations. 
We indeed observe the formation of a cusp at the reconnection point $s_r$ in all cases. 
%This is also illustrated by the colouring of the knot used in Fig.\ref{Fig:Unic}.b and the movie in the SI. 
Note that strictly speaking, no universal function of the curvature is observed. 
This is actually expected from the calculations of the curvature \eqref{eq:curvSelfSim} which shows a dependence on the values $A^+/A^-$ that differ from case to case. 
However, \eqref{eq:curvSelfSim} suggests that if the variable $\sqrt{1+(A^+/A^-)^2}(s-s_r)\kappa_{\rm max}$ is used instead, a universal form should be recovered. 
As shown in Fig.\ref{Fig:Cusps}.b the data indeed collapse into one universal function when using this new variable. 
The theoretical prediction \eqref{eq:curvSelfSim} is also plotted with dashed black line to appreciate the remarkable agreement. 
%The good agreement is remarkable. Note that there is no adjustable parameter (values of $A^\pm$ have been measured).

\begin{figure}
\centering
\includegraphics[width=0.95\linewidth]{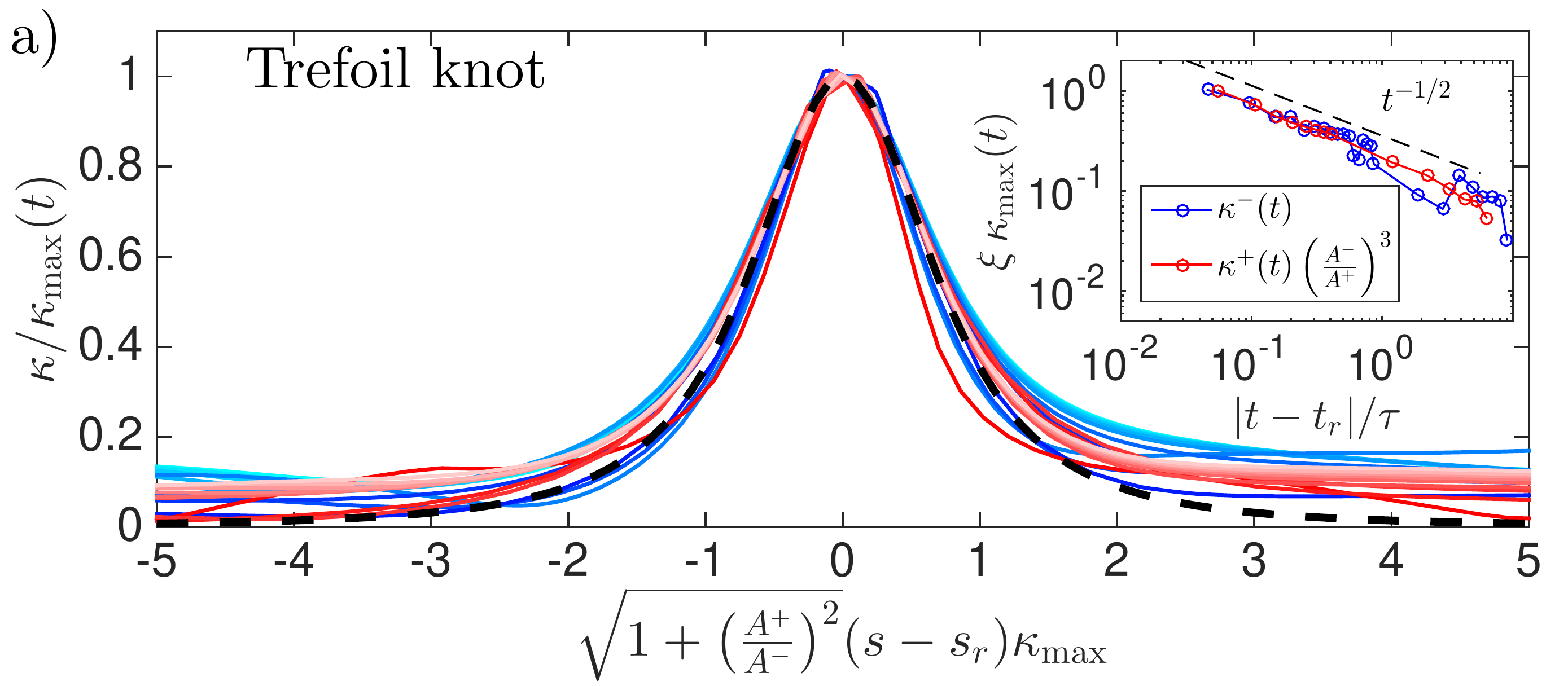} 
\includegraphics[width=0.95\linewidth]{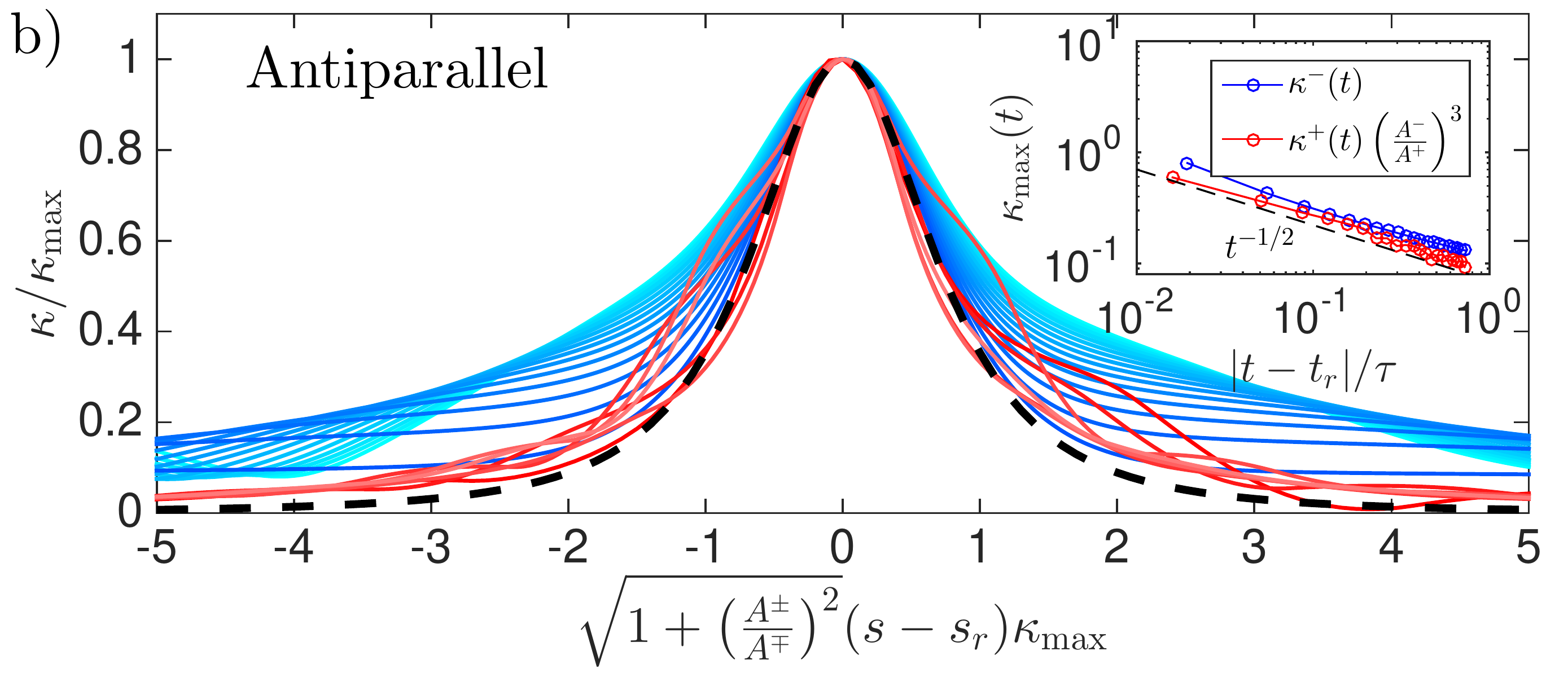} 
\caption{(Color online)  
a) Self-similar evolution of the curvature close to the reconnection point for the trefoil vortex. Blue lines (from light to dark) correspond to times before reconnection and red lines (from dark to light) to times after reconnection. The inset displays the temporal evolution of the maximum value of the curvature in LogLog scales before and after, normalized as suggested in \eqref{Eq:kappaMax}.
%\AD{b) 
b) Same plot as in a) but for the reconnection occurring in the antiparallel case.
In both the figures we consider times such that $|t-t_r|<0.5\tau$ and the dashed line shows the $t^{-1/2}$-scaling. In both figures the black dashed line displays the theoretical prediction (\ref{eq:curvSelfSim}).
 }
\label{Fig:CuspsSlefSimilar}
\end{figure}

We now study the temporal evolution of the curvature to determine if a self-similar evolution is observed. 
Figure \ref{Fig:CuspsSlefSimilar}a shows how the trefoil knot curvature, rescaled by their maximum values, almost perfectly collapse into one single plot, demonstrating the self-similar behavior for this configuration.
In the inset we plot the maximum value of the curvature as a function of time in LogLog scale. The predicted $t^{-1/2}$-scaling of \eqref{Eq:kappaMax} is clearly observed. 
%The self-similar evolution observed for the knot is actually a non-generic property in vortex reconnections. 
In Fig.\ref{Fig:CuspsSlefSimilar}b, we present the same analysis done for the antiparallel case where a clear break-down of the self-similar behavior is observed.
This can be explained by assuming a non-negligible value of $ \gamma $, hence a strong torsion, that breaks the validity of the expansion done to obtain \eqref{eq:curvSelfSim}, only recovered when times are very close to $ t_r $ as evident when comparing with the theoretical prediction displayed in dashed black line.
The temporal evolution of the maximum of curvature, shown in the inset of Fig.\ref{Fig:Cusps}b, still confirms the relations presented in \eqref{Eq:kappaMax}, namely the scaling $t^{-1/2}$-scaling normalized by the ratio of the pre-factors are confirmed. 
Note that the agreement is very good given the large value $(A^+/A^-)^3=4.15^3$.
For all other cases except tangle 2, self-similarity is observed (data not shown). 

\begin{figure}
\centering
\includegraphics[width=0.95\linewidth]{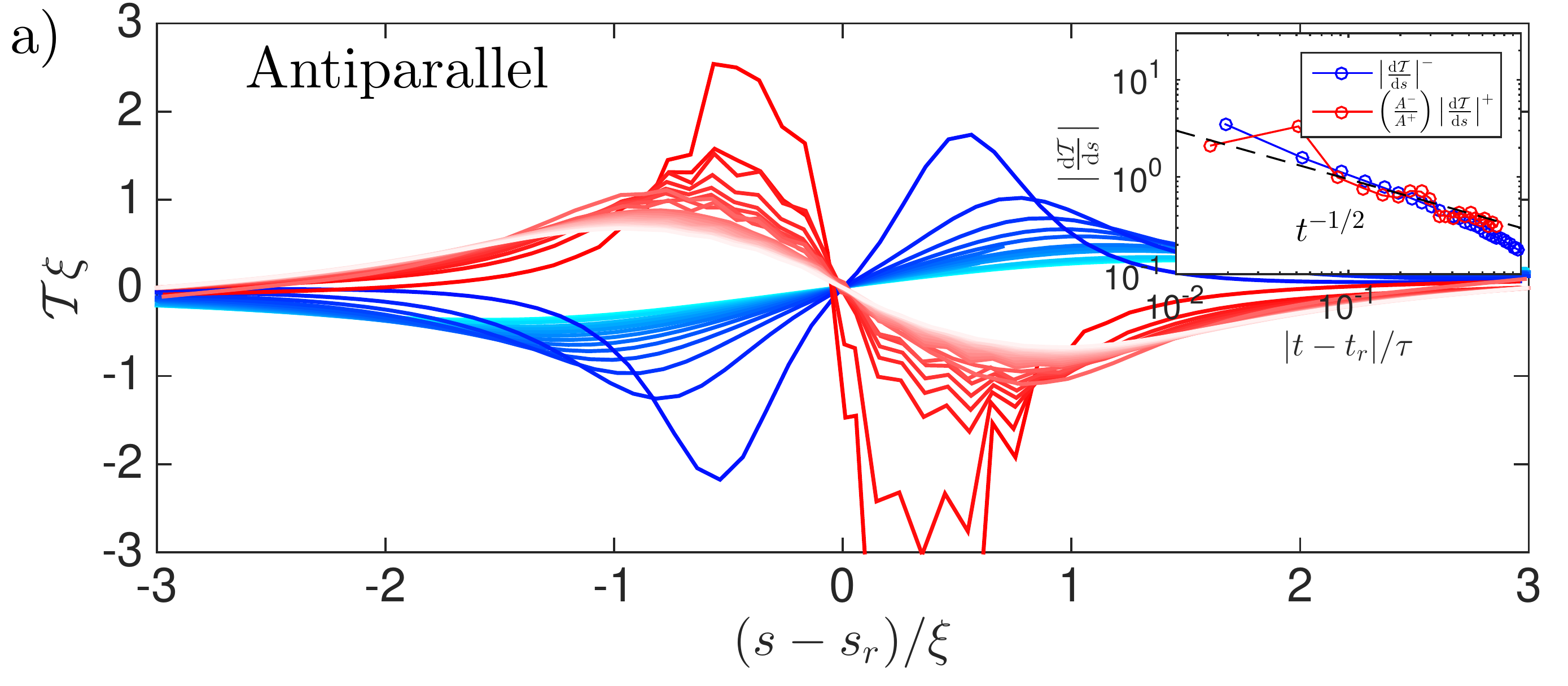}
\includegraphics[width=0.95\linewidth]{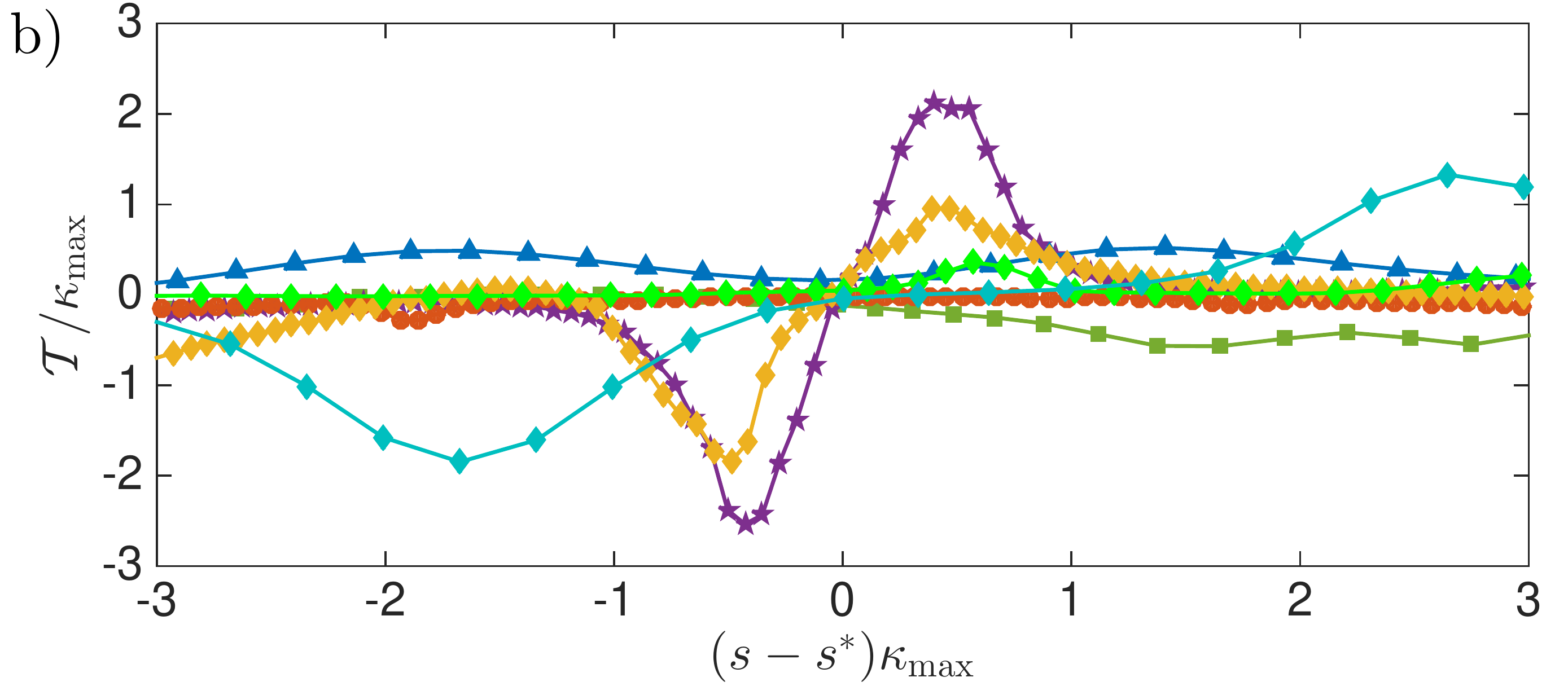}
\caption{(Color online)  
a) Temporal evolution of the torsion $\mathcal{T}$ as a function of the arc-length in the antiparallel case. Blue lines (from light to dark) correspond to times before reconnection and red lines (from dark to light) to times after reconnection. The inset displays the temporal evolution of the slope of torsion computed at the reconnection point $ s_r $ in LogLog scales before and after, normalized as $ (A^+/A^-) $.
b) Torsion $\mathcal{T}$ as a function of the arc-length close to the reconnection (just before) reconnection for all configurations (same legend as in Fig.\ref{Fig:Cusps}).
}
\label{Fig:Torsion}
\end{figure}

The breakdown of self-similarity is  predicted by \eqref{eq:curvSelfSim} when $A^+/A^-\neq 1$ and $\gamma\neq0$. 
A non-zero value of $\gamma$ is related, as we have seen, to torsion close to the reconnection point and a shock-like structure formation (see App~\ref{SubSec:LinearApprox}). 
In Fig.\ref{Fig:Torsion}.a we show the temporal evolution of the torsion $ \mathcal{T} $ for the antiparallel case.
The shock-like structure formation, and the linear behavior close to the reconnection point is clearly visible, thus explaining the breakdown of the self-similarity in Fig.\ref{Fig:CuspsSlefSimilar}.b.
The inset shows that the temporal evolution of the slope of the torsion at $ s_r $ obeys the scaling $ |t-t_r|^{-1/2} $ with the correct normalization $A^+/A^-$ suggested by the analytical calculations.
For completeness, in Fig.\ref{Fig:Torsion}.b we show the torsion normalized by the maximum value of the curvature for all the configurations close to the reconnection time. 
In all the other cases, except for tangle 2, the slope of torsion is almost zero at the reconnection point. 
Remarkably, the tangle 2 and antiparallel configurations correspond to the cases where vortices separate much faster than they approach (see Fig.\ref{Fig:Unic}.a).

We remark finally that measuring quantities such as curvature and torsion is numerically very challenging as they involve high-order derivatives.

\section*{Discussion\label{Sec:Conclu}}
The reconnection of quantized vortex filaments within the Gross--Pitaevskii model displays both universal and non-universal phenomena.
We found that close to the reconnection the approach and separation rates follow the same scaling $ \delta \sim (\Gamma t)^{1/2} $ and the vortex filaments always become locally anti-parallel. 
Previous numerical studies reported scaling rates in the form of power-law with exponents depending on the configuration. 
By dimensional analysis, any scaling different from $\alpha=1/2$ would introduce necessarily a new time or length-scale to the problem that needs to be made explicit.
The discrepancies in previous studies might be due to the fact that: (i) the computational domain is not big enough hence introducing a non-negligible system size length-scale, or (ii) the initial condition contains a considerable amount of sound waves such that the ${\it rms}$ value of the compressible kinetic energy can be used to construct an extra time-scale, or (iii) the observed scaling correspond to dynamical regimes occurring much farther/later than the reconnection event and thus driven by the specific vortex configuration and therefore non-universal. 
In that spirit, reconnections within Navier--Srokes flows, modified version of the model GP with non-local potential and/or high-order non-linearities to better replicate superfluid liquid helium, or coupled GP equations modeling multi-component or spinorial BECs could indeed lead to different scalings.

Our findings demonstrate that the pre-factors $ A^{\pm} $ are not universal in GP. 
However, once measured case by case, their ratio determines many properties of the reconnection dynamics.
Note that the easiest way to determine this ratio is to look at the medium/large scale reconnection angle $ \phi $ between the hyperbola asymptotes which should be an accessible quantity in superfluid experiments \cite{BewleyReconnectionPNAS, PhysRevLett.115.170402, serafini2016vortex}.
Let us also remark that the $t^{1/2}$-scaling we observed extends beyond the distance $\xi$. 
This suggests that the linear approximation might be used as a matching theory in order to relate measurements done well before and far from the reconnection events. 
BEC experimentalists are able today to study vortex dynamics and reconnections \cite{PhysRevLett.115.170402, serafini2016vortex}.
Our predictions should directly apply to those systems.

Finally, let us underline that understanding the dynamics of the reconnection events is crucial to provide a full comprehension of the dissipative processes occurring in superfluids in the low-temperature limit. 
It is largely believed that Kelvin waves play a fundamental role carrying the energy to the smallest scales where it gets finally dissipated by sound radiation. 
The cusps arising in the vortex filaments due to reconnection events are responsible for a rapid and efficient excitation of Kelvin waves at all scales. 
Here we provided an analytical formula for the dynamical formation of the cusps and we aim to use this result in further theoretical studies to estimate the  rate of radiation during reconnection.
Also, we have shown that non-negligible torsion of the reconnecting filaments implies the break-down of self-similarity, results in the formation of shock-like structures of the torsion.
This phenomenon seems to be linked to the large difference observed in the $A^\pm$ pre-factors, hence to extremes events where vortices separate much faster than they approach, and to the irreversibility of the reconnection events. 
We do not have yet a theoretical understanding of this fact and more data would be desirable to perform a detailed statistical analysis.  

\begin{acknowledgments}  
The authors were supported by the Royal Society and the Centre National de la Recherche Scientifique (CNRS) through the International Exchanges Cost Share Scheme (Ref. IE150527). 
Computations were carried out on the M{\'e}socentre SIGAMM hosted at the Observatoire de la C{\^o}te d'Azur and on the High Performance Computing Cluster supported by the Research and Specialist Computing Support service at the University of East Anglia.

\end{acknowledgments}

\appendix
\section{The Gross--Pitaevskii equation} \label{SubSec:GP}

In the limit of very low temperature a weakly-interacting Bose gas can be described using a mean field approximation in terms of a complex order parameter (or condensate wave function) $\psi$. Such system is governed by a dispersive non-linear wave equation called Gross--Piteaveskii equation
\begin{equation}
i\hbar\dertt{\psi} =- \alps \gd \psi + \bet|\psi|^2\psi -\mu\psi,
\label{Eq:GPE}
\end{equation}
where $m$ is the mass of the bosons and $g=4 \pi a \hbar^2 / m$, with $a$ the boson $s$-wave scattering length. 
The chemical $\mu$ can be in principle absorbed by a global phase shift. 
Although formally derived for BECs, the GP model qualitatively reproduces many aspects of superfluid liquid Helium too. 
It can be used to model classical vortex dynamics in situations where a large scale separation between the vortex core and the size of such vortex is present.

The GP equation posses a Hamiltonian structure and conserves the total number of particles 
\begin{equation}
 N=\int|\psi|^2\mathrm{d}^3{\bf x},
\end{equation}
the total energy
\begin{equation}\label{E}
H= \int \left( \frac{\hbar^2}{2m} |\nabla\psi|^2 + \frac{g}{2} |\psi|^4 \right) \mathrm{d}^3{\bf x} \, ,
 \end{equation}
and the total momentum
\begin{equation}\label{eq:Lin_mom}
{\bf P}=\frac{\hbar}{2i}\int[\psi^*\nabla\psi-\psi\nabla\psi^*]\mathrm{d}^3{\bf x} \, .
\end{equation} 
The speed of sound for such model is given by $ c=\sqrt{g \rho_0}/m $. This value can be derived by linearizing \eqref{Eq:GPE} about a constant value $\psi= \sqrt{\rho_0/m}=\sqrt{\mu/g}$. It is also possible to identify a characteristic length $\xi= \hbar / \sqrt{2 \rho_0 g} $, called healing length, representing the scale where the linear contribution in \eqref{Eq:GPE} equals the non-linear one.  Dispersive effects will then take place for length scales smaller than $\xi$.
\eqref{Eq:GPE} can be rewritten in term of the two physical quantities $c$ and $\xi$ as 
\begin{equation}
i\dertt{\psi} = \frac{c}{\sqrt{2}\xi}\left(-\xi^2\nabla^2\psi-\psi+\frac{m}{\rho_0}|\psi|^2\psi \right) \, .
\label{Eq:GPhydroApp}
\end{equation}
By using the Madelung's transformation
\begin{equation}\label{eq:Madelung}
\psi({\bf x},t)=\sqrt{\frac{\rho({\bf x},t)}{m}}e^{i \frac{\varphi({\bf x},t)}{\sqrt{2}c\xi} },
\end{equation}
it possible to relate the order parameter $\psi$ to a compressible, irrotational and barotropic superfluid having density $\rho({\bf x},t)$ and velocity ${\bf v}={\bf \nabla} \varphi$. 
Indeed, plugging transformation (\ref{eq:Madelung}) in \eqref{Eq:GPhydro} we directly obtain
\begin{eqnarray}
\dertt{\rho}+\nabla\cdot \left( \rho {\bf v} \right)& = &0\label{Eq:Cont}\\
\dertt{\phi}+\frac{1}{2}{\bf v}^2 &= & c^2\frac{(\rho_0-\rho)}{\rho_0}+c^2\xi^2\frac{\nabla^2\sqrt{\rho}}{\sqrt{\rho}}.\label{Eq:Bernouilli}
\end{eqnarray}
\eqref{Eq:Cont} and \eqref{Eq:Bernouilli} are the continuity equation and the Bernoulli equation respectively, 
except for the last term in \eqref{Eq:Bernouilli} which is called quantum pressure and has no analog in classical fluid mechanics. 

Although the velocity field defined by the Madelung transformation \eqref{eq:Madelung} is potential, solutions with non-zero circulation can exist in the form of topological defects of the order parameter $\psi$. 
For such vortex solutions the vorticity is supported on the curves (nodal lines) where the density field vanishes and the phase is not defined. In order to ensure that the order parameter stays single-valued, the circulation around such nodal lines must be constant and equal to a multiple of the Onsager-Feynman quantum of circulation $\Gamma=h/m=2\sqrt{2}\pi c\,\xi$.
For this reason, nodal lines of the order parameter are called quantum (or quantized) vortices.
The region around the topological defect where the density drops to zero is called vortex core and its size is of the order of the healing length $\xi$. 
%Such vortices have a density depletion core which size is order of the coherence length $\xi$. 
%As two conditions are required for the wavefunction to vanish, 
%Quantum vortices generally correspond to points in 2D and filaments in 3D.
The hydrodynamical interpretation of superfluids is thus the one of a compressible (dispersive) flow where vorticity is a distribution (a superposition of Dirac deltas) supported on the vortex filament. 

We would like to remark that often quantum vortices are misleadingly referred as the \emph{singularities} of the system, as the velocity field diverges as $1/r$ where $r$ is the distance to the filament. 
This divergence is just a consequence of the change of coordinates given by the Madelung transformation. 
At the vortex position, the order parameter solution of the GP equation is a smooth field. 
We can thus precisely track vortices finding the zeros of $ \psi $ as described in the next section.

\section{Vortex Tracking Algorithm\label{SubSec:Tracking}}
\begin{figure}
 \centering
      \includegraphics[scale=0.35]{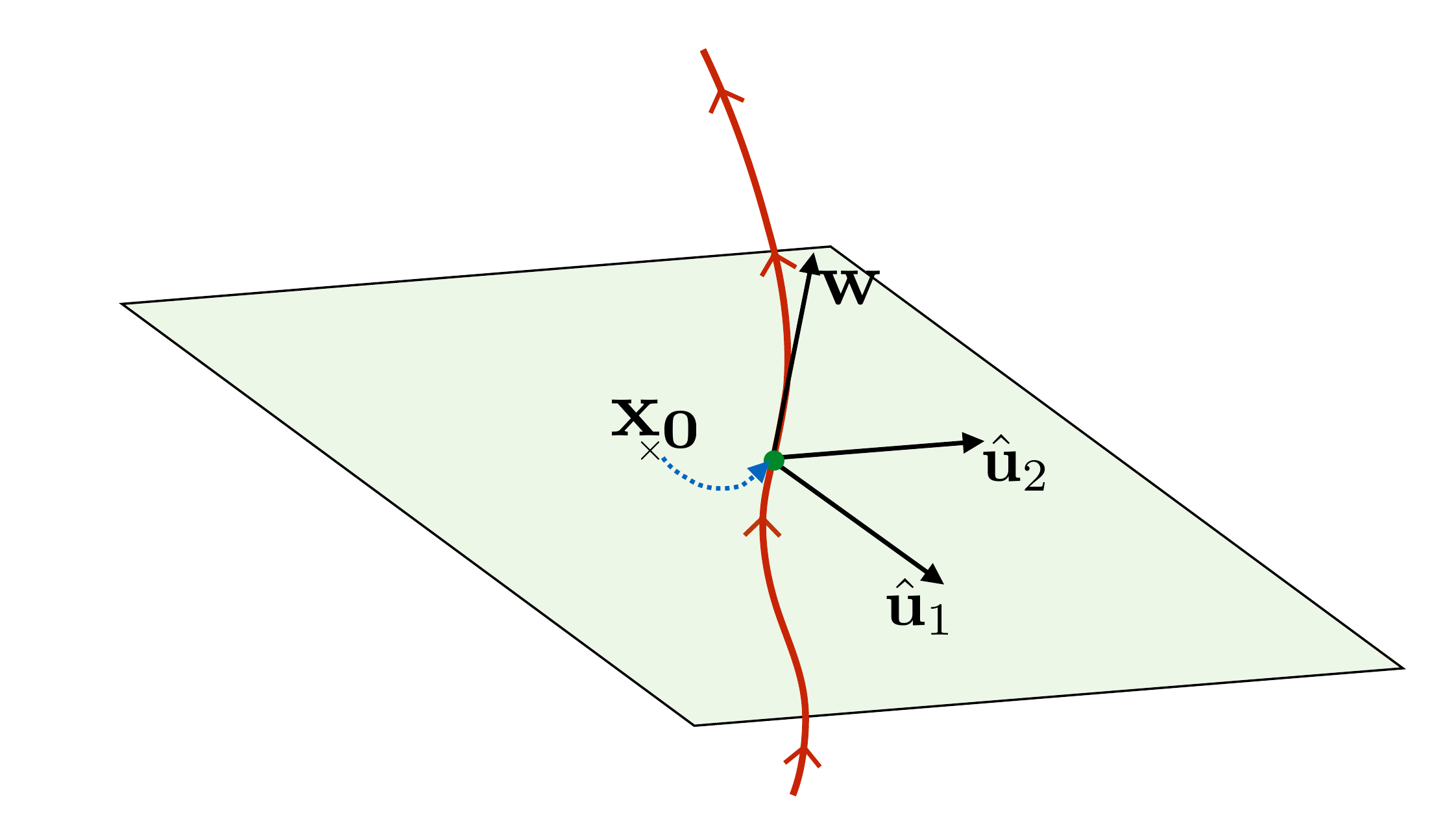}
 \caption{Sketch of the plane on which the Newton--Raphson method is implemented.
\label{fig:Sketch}}
\end{figure}

We have recently developed a robust and accurate algorithm to track vortex lines of the order parameter $\psi$ in arbitrary geometries. 
The details of the algorithm and accuracy of the method can be found in \cite{VilloisTrackingAlgo}.
We recall here the basic ideas.
A quantized vortex line in three dimensions corresponds to a nodal line defined by 
\begin{equation}
Re[\psi(x,y,z)]=Im[\psi(x,y,z)]=0 \, .
\end{equation}
The algorithm is based on a Newton--Raphson method to find zeros of $\psi$ and on the knowledge of the \emph{pseudo-vorticity} field ${\bf W}=\nabla Re[\psi]\times \nabla Im[\psi]$, always tangent to the filaments, to follow vortex lines \cite{2014arXiv1410.1259R}. 
Starting from a point ${\bf x_0}$ where the density $|\psi|^2$ is below a given small threshold (therefore very close to a vortex), we define the orthogonal plane to the vortex line using ${\bf W}({\bf x_0})$. 
The plane is then spanned by the two directors $\hat{{\bf u}}_1$ and $\hat{{\bf u}}_2$ as illustrated in Fig.\ref{fig:Sketch}.
A better approximation for the vortex position ${\bf x_{\rm v}}$ on the plane 
is then given by ${\bf x_1=x_0+\delta x}$. Here the increment ${\bf \delta x}$ is obtained using the Newton--Raphson formula (the linear approximation):
\begin{equation}
0=\psi({\bf x_0+\delta x})\approx\psi({\bf x_0})+ J({\bf x_0}) {\bf \delta x} \, ,
\end{equation}
where $J({\bf x_0})$ is the Jacobian matrix expressed as
\begin{equation}\label{Jac_gen}
J=\begin{pmatrix}
 \nabla Re[\psi]\cdot \hat{{\bf u}}_1  &   \nabla Re[\psi]\cdot \hat{{\bf u}}_2 \\
 \nabla Im[\psi]\cdot \hat{{\bf u}}_1&  \nabla Im[\psi]\cdot \hat{{\bf u}}_2
  \end{pmatrix}.
\end{equation}
The increment can be therefore calculated using  ${\bf \delta x}= - J^{-1}({\bf x_0})\cdot  (Re[\psi({\bf x_0})],Im[\psi({\bf x_0})])^T$. 
Sufficiently close to the line the Jacobian matrix is always a non-singular $2\times2$ matrix so its inverse can be computed. 
We underline that the method requires the evaluation of the Jacobian~\eqref{Jac_gen} at intermesh points. Making use of the spectral representation of $\psi$, we can precisely compute those values using Fourier transforms. 
This process can be iterated until the exact location ${\bf x_{\rm v}}$ is determined upon a selected convergence precision. 

To track the following vortex point of the same line we use as a next initial guess ${\bf x_0}={\bf x_{\rm v}}+\zeta {\bf W}$, which is obtained evolving along $ {\bf W} $  by a small step $\zeta$.
The process is reiterated until the entire line is tracked and closed, then repeated with another line until the whole computation domain has been fully explored. 

\section{Linear approximation detailed calculations\label{SubSec:LinearApprox}}

\begin{figure}[h]
 \centering
      \includegraphics[scale=0.25]{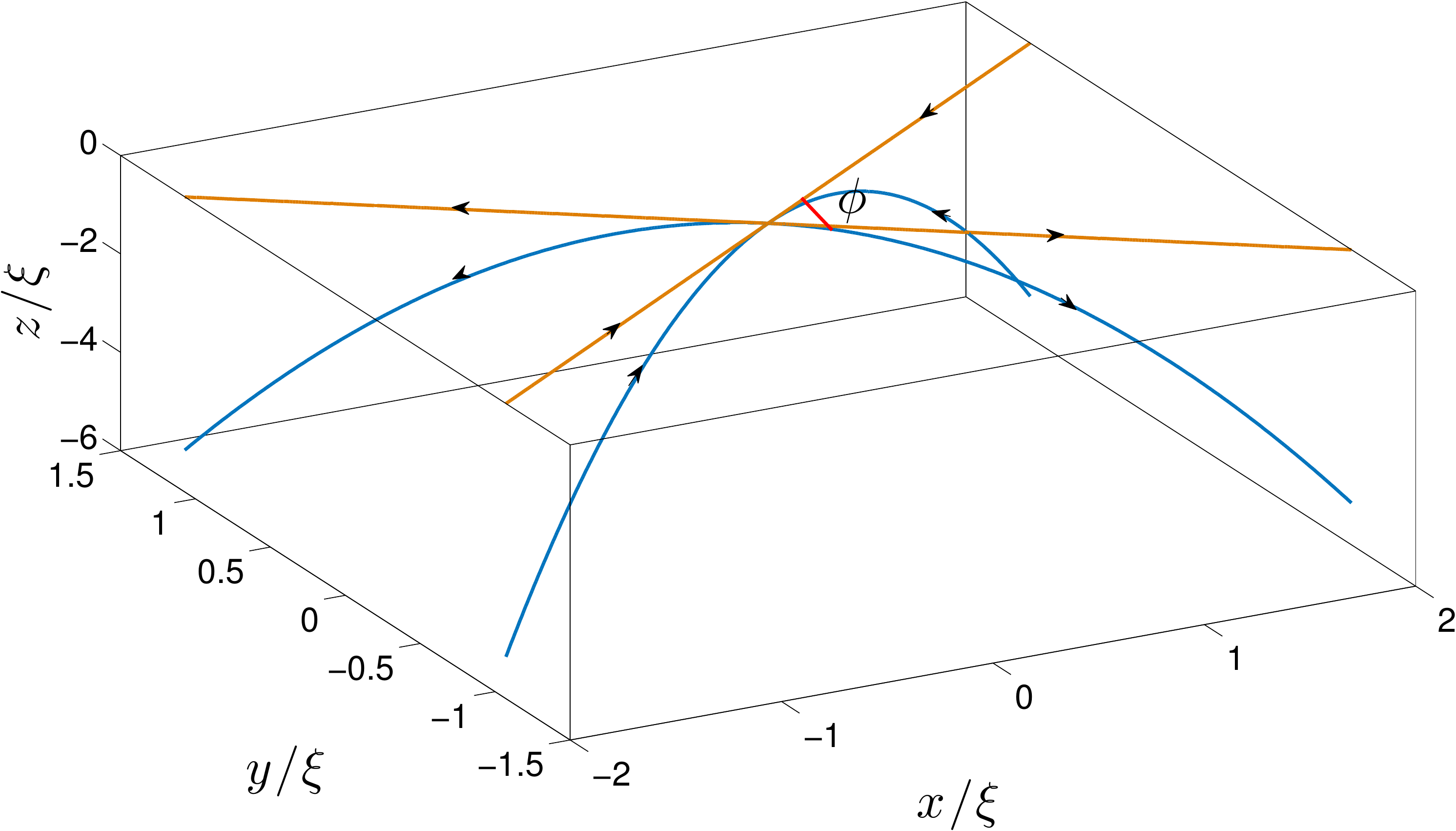}
 \caption{Plot of the initial condition using $\gamma=0.01$, $\beta=1/2$ and $a=1$. The vortex filaments are shown in blue while their projection on the $x$-$y$ is shown in orange.The arrows identify the circulation around each vortex. 
\label{fig:in_cond}}
\end{figure}

A first analytical study of a reconnection event in the GP model have been provided by Nazarenko and West \cite{nazarenko2003analytical}, where it is shown that two vortices are anti-parallel during a reconnection and their distance scales as $\delta(t)\sim t^{1/2}$. 
In the same spirit of the work \cite{nazarenko2003analytical}, we assume that inside the vortex core the non-linear term of the GP equation can be neglected and so a reconnection event should be governed by the (linear) Schr{\"o}dinger equation. For the sake of simplicity, in dimensionless units this equations reads
\begin{equation}\label{SCHR}
i\partial_{t}\psi+\frac{1}{2}\nabla^2\psi=0.
\end{equation}
Note that we absorbed the parameters $c$ and $\xi$ in \eqref{Eq:GPhydro} by a suitable time and space rescaling. %In this work we decided expressed all the lengths in units of the healing length $\xi$ and its characteristic time $\tau=\xi/c$. 
We remark that in \cite{nazarenko2003analytical} reconnections are studied just on a plane, whereas here we consider vortex filaments with non-zero torsion.
At the reconnection time $t_r$ we use as initial condition the ansatz
\begin{equation}
\psi_r(x,y,z)=z +\frac{\gamma}{a}(x^2+y^2)+i (az+\beta x^2-y^2).
\end{equation}
Looking for $\psi_r=0$ one can recover the vortex profile, given  by the curves
\begin{equation}\label{curve}
\mathbf{R}(s)=\left(s,\,\ \pm s \sqrt{\frac{\beta-\gamma}{\gamma+1}},\,\ -s^2\frac{\gamma(\beta+1)}{a(\gamma+1)}\right),
\end{equation}
where $s$ is the parametrization of the curve. 
%
%\begin{equation}
%z=-\frac{\gamma(x^2+y^2)}{a}
%\end{equation}
%and 
%\begin{equation}\label{hyper}
%y=\pm x\sqrt{\frac{\beta - \gamma}{\gamma+1}}.
%\end{equation}
%
We note that \eqref{curve} requires that
\begin{equation}\label{first_cond}
\frac{\beta - \gamma}{\gamma+1}>0.
\end{equation}
In figure \ref{fig:in_cond} we plot the the vortex filaments $\mathbf{R}(s)$ (blue lines) for our initial condition.
 The vortices projected on the $x$-$y$ planes form two hyperbola (orange lines) crossing at the reconnection point. We note that the values $\beta$ and $\gamma$ fix the angle 
 \begin{equation}\label{angle}
\phi=2 \tan^{-1}\left(\sqrt{\frac{1+\gamma}{\beta-\gamma}}\right)
\end{equation} 
 between the two hyperbola. The arrows identify the circulation around each vortex. 

\begin{figure}[h]
 \centering
      \includegraphics[scale=0.26]{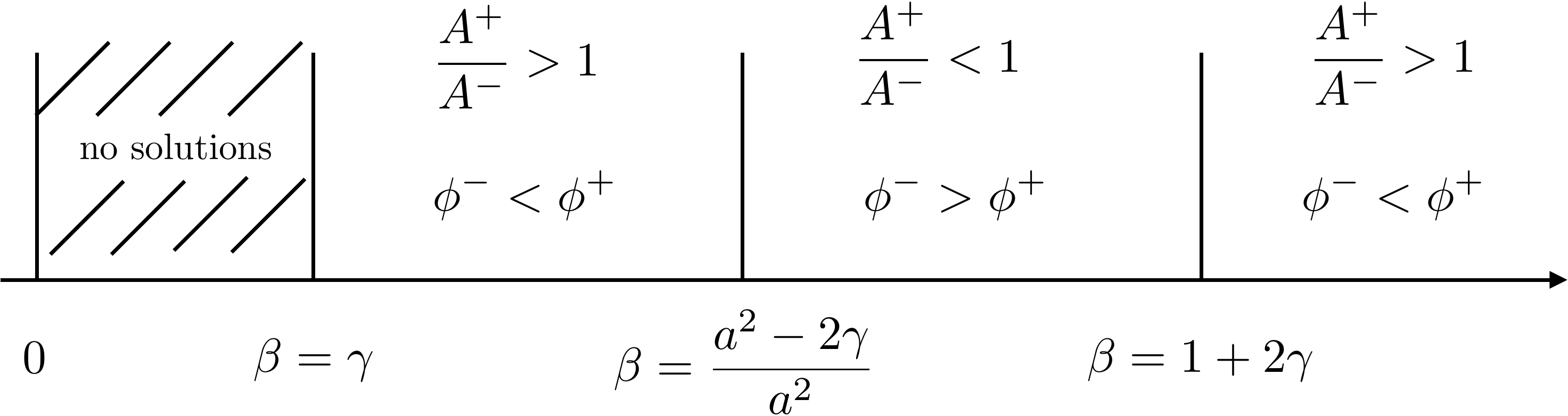}
 \caption{Dependence of the ratio $\frac{A^{+}}{A^{-}}$ and the angles $\phi^{-}$ and $\phi^{+}$ on different values of $\beta$.
\label{fig:graph}}
\end{figure}

The formal solution of equation~\eqref{SCHR} is given by 
\begin{equation}
\psi(t) = e^{i \frac{1}{2}(t-t_r)\nabla^2}\psi_r 
\end{equation}
where $t_r$ is the time when the reconnection occurs.
The choice of a second order polynomial for $\psi_r$ allows us to find the exact solution of \eqref{SCHR}:
\small  
\begin{equation}
\psi(t) =z +\frac{\gamma}{a}(x^2+y^2) - 2t (\beta-1)+i\left(az+\beta x^2-y^2 + 4(t-t_r)\frac{\gamma}{a}
\right).
\end{equation}
\normalsize
Assuming $a> 0$ and $\gamma< \beta < \frac{a^2-2\gamma}{a^2} $ the vortex lines before the reconnection ($t<t_r$) are given by
\small
\begin{equation}
\begin{split}
\mathbf{R}_{1,2}^{-}(s,t)=
& (s,\,\ \pm \sqrt{ \frac{(t_r-t)(a^2(1-\beta )-2 \gamma)+as^2(\beta-\gamma)}{a(\gamma+1)}},\\
&\frac{(t-t_r)(a^2(\beta-1)-2\gamma^2)-a\gamma(\beta+1)s^2}{(\gamma+1)a^2} )
\end{split}
\end{equation}
\normalsize
while after the reconnection ($t>t_r$)
\small
\begin{equation}
\begin{split}
\mathbf{R}_{1,2}^{+}(s,t)=
&(\pm \sqrt{ \frac{(t-t_r)(a^2(1-\beta)-2\gamma)+as^2(1+\gamma)}{a(\beta -\gamma)}},\,\ s ,\\
& \frac{(t-t_r)(a^2(\beta-1)+2\gamma^2)-a\gamma(\beta+1)s^2}{(\beta-\gamma)a^2} ).
\end{split}
\end{equation}
\normalsize
From the above curves we observe that the two vortices approach along the $y$-direction and separate along the $x$-direction. It follows that
\begin{equation}
\delta^{\pm}(t)=|\mathbf{R}_{1}^{\pm}(0,t)-\mathbf{R}_{2}^{\pm}(0,t)|=\sqrt{2\pi} A^{\pm}|t-t_r|^{1/2},
\end{equation} 
where the ratio of pre-factors satisfies 
\begin{equation}
\frac{A^{+}}{A^{-}}=\sqrt{\frac{1+\gamma}{\beta-\gamma}}>1.
\end{equation}
From equation~\eqref{angle} we can see how the quantity $\frac{A^{+}}{A^{-}}$ is related to the angle $\phi$.
Calling $\phi^{-}$ the angle of the approaching vortices and $\phi^{+}$ the angle of the separating vortices, we can conclude that for $\beta<\frac{a^2-2\gamma}{a^2}$ then $\phi^{-}>\phi^{+}$. On the other hand, when  $\beta>\frac{a^2-2\gamma}{a^2}$ the two vortices approach along the $x$-direction and separate along the $y$-direction with $\frac{A^{+}}{A^{-}}=\sqrt{\frac{\beta-\gamma}{1+\gamma}}$. For sake of completeness in Fig.\ref{fig:graph} we show the values of the ratio $\frac{A^{+}}{A^{-}}$ and the angles $\phi^{-}$ and $\phi^{+}$ for different values of $\beta$. We note that $\frac{A^{+}}{A^{-}}<1$ for $\frac{a^2-2\gamma}{a^2}<\beta<1+2\gamma$ while  $\frac{A^{+}}{A^{-}}>1$ for $\beta>1+2\gamma$. 
As a final remark, we note that changing the sign of $a$ corresponds to look at the reconnection back in time, hence each value of $\frac{A^{+}}{A^{-}}$ in Fig.\ref{fig:graph} will then be reversed.\\

The linear approximation also allows for computing the curvature 
\begin{equation}
\kappa(s,t)=\frac{|\mathbf{R}'(s,t)\times \mathbf{R}''(s,t)|}{|\mathbf{R}'(s,t)|^3}
\end{equation}
and torsion 
\begin{equation}
\mathcal{T}(s,t)=\frac{(\mathbf{R}'(s,t)\times \mathbf{R}''(s,t))\cdot  \mathbf{R}'''(s,t)}{|\mathbf{R}'(s,t)\times \mathbf{R}''(s,t)|^2}
\end{equation}
of the vortex lines.

%\small
%\begin{equation}
%\begin{split}
%\kappa^-_{\rm max}(t)=
%&\kappa^-(0, t) =\\
%&\sqrt{\frac{4\gamma^2(1+\beta)^2(a^2(\beta-1)+2\gamma)(t-t_r)+a^3(\beta-\gamma)^2(1+\gamma)}{a^2  (1+\gamma)^2 (a^2 (\beta-1)+2 \gamma)(t-t_r)}}
%\end{split}
%\end{equation}
%\normalsize
%and 
%\small
%  \begin{equation}
%\begin{split}
%\kappa^+_{\rm max}(t)=
%&\kappa^+(0, t) =\\
%&\sqrt{\frac{4\gamma^2(1+\beta)^2(a^2(\beta-1)+2\gamma)(t-t_r)-a^3(\beta-\gamma)^2(1+\gamma)^2}{a^2  (\beta-\gamma) (a^2 (\beta-1)+2 \gamma)(t-t_r)}}
%\end{split}.
%\end{equation}
%\normalsize
The curvature can be directly evaluated. Its maxima as a function of time before and after reconnection are given by
\footnotesize
\begin{equation}
\kappa^-_{\rm max}(t)\sqrt{\frac{4\gamma^2(1+\beta)^2(a^2(\beta-1)+2\gamma)(t-t_r)+a^3(\beta-\gamma)^2(1+\gamma)}{a^2  (1+\gamma)^2 (a^2 (\beta-1)+2 \gamma)(t-t_r)}}
\end{equation}
\normalsize
and
\footnotesize
  \begin{equation}
\kappa^+_{\rm max}(t)=\sqrt{\frac{4\gamma^2(1+\beta)^2(a^2(\beta-1)+2\gamma)(t-t_r)-a^3(\beta-\gamma)^2(1+\gamma)^2}{a^2  (\beta-\gamma) (a^2 (\beta-1)+2 \gamma)(t-t_r)}}
\end{equation}
\normalsize
respectively. The present calculation predicts $ \kappa^\pm_{\max}(t) \propto |t-t_{\rm r} |^{-1/2}$ that also corresponds to a dimensional analysis prediction. In addition, the linear approximation predicts that  $ \kappa^+_{\max}/ \kappa^-_{\max}=(A^+/A^-)^{3}$ in the limit of $t\rightarrow t_r$. This non-trivial result can not be found by dimensional arguments.
 Moreover one can show that
$ \kappa^\pm $ presents a self-similar behavior close to the reconnection point of the form $ \kappa^\pm(s, t) =\kappa^\pm_{\rm max}(t)\Phi^\pm(\zeta_\pm)$,  where $\zeta^\pm=(s-s_r)\kappa^\pm_{\rm max}(t)$ and $s_r$ is the coordinate of the reconnecting point.  For small values of $\gamma$, these self-similar functions can be found to be
\begin{equation}
\begin{split}
\Phi^\pm(\zeta)
&=\frac{1\pm \frac{3}{2} \frac{ (\beta^{\pm1} +1) \zeta ^2}{1+(\beta^{\pm1} +1) \zeta ^2}\gamma }{\left(1+(\beta^{\pm1} +1) \zeta^2\right)^{3/2}} +O(\gamma^2)  \\
&=\frac{1}{\left[ 1+\left(\left(\frac{A^{\mp}}{A^{\pm}}\right)^2+1\right)\xi^2\right]^{3/2}}+O\left(\eta^\pm\gamma^2 \frac{(t-t_r)}{\tau}\right),
\end{split}
\end{equation}
where $\eta^\pm=(A^{\mp}/A^{\pm})^2-1$. Remarkably, once the ratio $A^+/A^-$ is reintroduced, $\gamma$ only appears as quadratic correction to the self-similar form. Note that within this approximation, self-similarity is destroyed when $\eta^\pm\gamma^2 (t-t_r)/\tau$ is of order $1$.

We note that in case one chooses $\beta>\frac{a^2-2\gamma}{a^2}$ then $\frac{A^{+}}{A^{-}}=\sqrt{\frac{\beta-\gamma}{1+\gamma}}$ and 
\begin{equation}
\left[\Phi^\pm(\zeta)\right]_{\beta>\frac{a^2-2\gamma}{a^2}}=\left[\Phi^\mp(\zeta)\right]_{\beta<\frac{a^2-2\gamma}{a^2}}.
\end{equation}
The former calculations were evaluated using a symbolic computation software. 

\begin{figure}[h]
 \centering
      \includegraphics[scale=0.32]{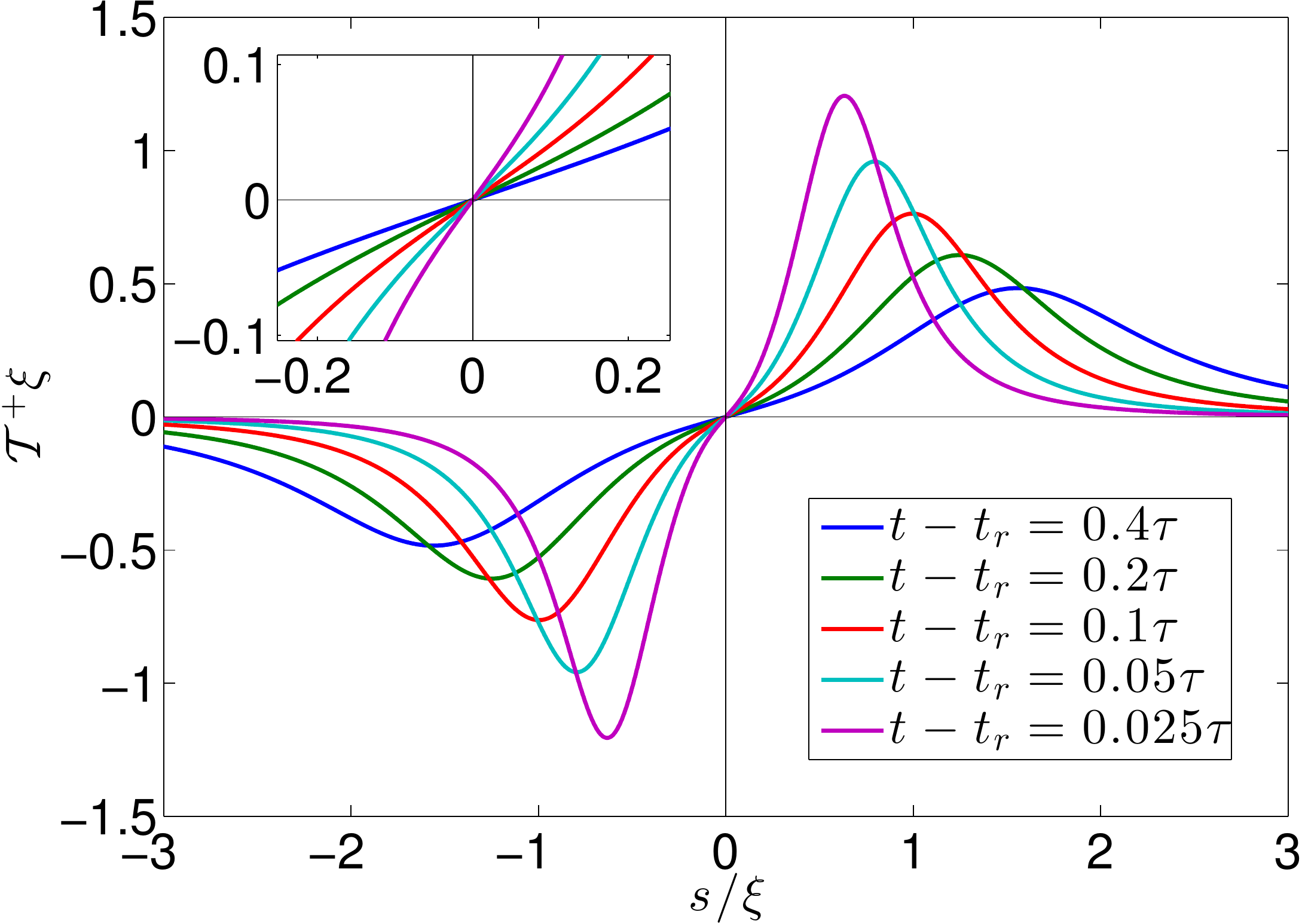}
 \caption{Plot of the torsion versus the $y$-coordinate, for different time steps using $\gamma=0.01$ , $\beta=1/2$ and $a=1$.
\label{fig:torsion}}
\end{figure}

Finally, the torsion $\mathcal{T}^\pm(s,t)$ of vortex line can be also computed within this approximation. It vanishes at $s_r$ (suggesting a locally planar reconnection), however it changes sign linearly at this point. Its slope is given by
\begin{equation}
\frac{\mathrm{d}\mathcal{T}^+}{\mathrm{d}s}=-\gamma\frac{3\sqrt{2} (1 + \beta)}{\sqrt{a (\beta - \gamma)}\sqrt{ (t-t_r) (a^2 (1-\beta) - 2 \gamma})},
\end{equation}
and it diverges as $\gamma|t-t_r|^{-1/2}$. The torsion thus develops shock-like structures as displayed in Fig.\ref{fig:torsion}.
The inset in Fig. \ref{fig:torsion} shows the linear behavior close to the reconnection point. It is possible to prove analytically that the ratio of the slopes is given by $\left.\frac{\mathrm{d}\mathcal{T}^+}{\mathrm{d}s}/\frac{\mathrm{d}\mathcal{T}^-}{\mathrm{d}s}\right|_{s=s_r}=A^+/A^-$. The full formulas for the torsion are too long to be presented here.

\bibliography{Reconnections}

%Merlin.mbs v4.21 2009-07-09.
\begin{thebibliography}{10}%
\makeatletter
\providecommand \@ifxundefined [1]{%
 \ifx #1\undefined \expandafter \@firstoftwo
 \else \expandafter \@secondoftwo
\fi
}%
\providecommand \@ifnum [1]{%
 \ifnum #1\expandafter \@firstoftwo
 \else \expandafter \@secondoftwo
\fi
}%
\providecommand \enquote [1]{``#1''}%
\providecommand \bibnamefont  [1]{#1}%
\providecommand \bibfnamefont [1]{#1}%
\providecommand \citenamefont [1]{#1}%
\providecommand\href[0]{\@sanitize\@href}%
\providecommand\@href[1]{\endgroup\@@startlink{#1}\endgroup\@@href}%
\providecommand\@@href[1]{#1\@@endlink}%
\providecommand \@sanitize [0]{\begingroup\catcode`\&12\catcode`\#12\relax}%
\@ifxundefined \pdfoutput {\@firstoftwo}{%
 \@ifnum{\z@=\pdfoutput}{\@firstoftwo}{\@secondoftwo}%
}{%
 \providecommand\@@startlink[1]{\leavevmode\special{html:<a href="#1">}}%
 \providecommand\@@endlink[0]{\special{html:</a>}}%
}{%
 \providecommand\@@startlink[1]{%
  \leavevmode
  \pdfstartlink
   attr{/Border[0 0 1 ]/H/I/C[0 1 1]}%
   user{/Subtype/Link/A<</Type/Action/S/URI/URI(#1)>>}%
  \relax
 }%
 \providecommand\@@endlink[0]{\pdfendlink}%
}%
\providecommand \url  [0]{\begingroup\@sanitize \@url }%
\providecommand \@url [1]{\endgroup\@href {#1}{\urlprefix}}%
\providecommand \urlprefix [0]{URL }%
\providecommand \Eprint[0]{\href }%
\@ifxundefined \urlstyle {%
  \providecommand \doi [1]{doi:\discretionary{}{}{}#1}%
}{%
  \providecommand \doi [0]{doi:\discretionary{}{}{}\begingroup
  \urlstyle{rm}\Url }%
}%
\providecommand \doibase [0]{http://dx.doi.org/}%
\providecommand \Doi[1]{\href{\doibase#1}}%
\providecommand \bibAnnote [3]{%
  \BibitemShut{#1}%
  \begin{quotation}\noindent
    \textsc{Key:}\ #2\\\textsc{Annotation:}\ #3%
  \end{quotation}%
}%
\providecommand \bibAnnoteFile [2]{%
  \IfFileExists{#2}{\bibAnnote {#1} {#2} {\input{#2}}}{}%
}%
\providecommand \typeout [0]{\immediate \write \m@ne }%
\providecommand \selectlanguage [0]{\@gobble}%
\providecommand \bibinfo [0]{\@secondoftwo}%
\providecommand \bibfield [0]{\@secondoftwo}%
\providecommand \translation [1]{[#1]}%
\providecommand \BibitemOpen[0]{}%
\providecommand \bibitemStop [0]{}%
\providecommand \bibitemNoStop [0]{.\EOS\space}%
\providecommand \EOS [0]{\spacefactor3000\relax}%
\providecommand \BibitemShut [1]{\csname bibitem#1\endcsname}%
%</preamble>
\bibitem{Priest1999}%
  \BibitemOpen
  \bibfield{author}{%
  \bibinfo {author} {\bibfnamefont{E.~R.}\ \bibnamefont{Priest}},\ }%
  \enquote{\bibinfo {title} {Heating the solar corona by magnetic
  reconnection},}\ in\ \Doi{10.1007/978-94-011-4203-8_8}{\emph{\bibinfo
  {booktitle} {Plasma Astrophysics And Space Physics: Proceedings of the VIIth
  International Conference held in Lindau, Germany, May 4--8, 1998}}},\
  \bibinfo {editor} {edited by\ \bibinfo {editor} {\bibfnamefont{J{\"o}rg}\
  \bibnamefont{B{\"u}chner}}, \bibinfo {editor} {\bibfnamefont{Ian}\
  \bibnamefont{Axford}}, \bibinfo {editor} {\bibfnamefont{Eckart}\
  \bibnamefont{Marsch}},\ and\ \bibinfo {editor} {\bibfnamefont{Vytenis}\
  \bibnamefont{Vasyli{\={u}}nas}}}\ (\bibinfo {publisher} {Springer
  Netherlands},\ \bibinfo {address} {Dordrecht},\ \bibinfo {year} {1999})\ pp.\
  \bibinfo {pages} {77--100},\ ISBN \bibinfo {isbn} {978-94-011-4203-8},\
  \url{http://dx.doi.org/10.1007/978-94-011-4203-8_8}%
  \bibAnnoteFile{NoStop}{Priest1999}%
\bibitem{Kida1994}%
  \BibitemOpen
  \bibfield{author}{%
  \bibinfo {author} {\bibfnamefont{S.}~\bibnamefont{Kida}}\ and\ \bibinfo
  {author} {\bibfnamefont{M.}~\bibnamefont{Takaoka}},\ }%
  \bibfield{title}{%
  \enquote{\bibinfo {title} {Vortex reconnection},}\ }%
  \bibfield{journal}{%
  \Doi{10.1146/annurev.fl.26.010194.001125}{\bibinfo {journal} {Annual Review
  of Fluid Mechanics}}\ }%
  \textbf{\bibinfo {volume} {26}},\ \bibinfo {pages} {169--177} (\bibinfo
  {year} {1994}),\
  \Eprint{http://arxiv.org/abs/http://dx.doi.org/10.1146/annurev.fl.26.010194.%
001125}{http://dx.doi.org/10.1146/annurev.fl.26.010194.001125},\
  \url{http://dx.doi.org/10.1146/annurev.fl.26.010194.001125}%
  \bibAnnoteFile{NoStop}{Kida1994}%
\bibitem{KoplikLevinPRL1993}%
  \BibitemOpen
  \bibfield{author}{%
  \bibinfo {author} {\bibfnamefont{Joel}\ \bibnamefont{Koplik}}\ and\ \bibinfo
  {author} {\bibfnamefont{Herbert}\ \bibnamefont{Levine}},\ }%
  \bibfield{title}{%
  \enquote{\bibinfo {title} {Vortex reconnection in superfluid helium},}\ }%
  \bibfield{journal}{%
  \Doi{10.1103/PhysRevLett.71.1375}{\bibinfo {journal} {Phys. Rev. Lett.}}\ }%
  \textbf{\bibinfo {volume} {71}},\ \bibinfo {pages} {1375--1378} (\bibinfo
  {month} {Aug}\ \bibinfo {year} {1993}),\
  \url{http://link.aps.org/doi/10.1103/PhysRevLett.71.1375}%
  \bibAnnoteFile{NoStop}{KoplikLevinPRL1993}%
\bibitem{Zhike2016}%
  \BibitemOpen
  \bibfield{author}{%
  \bibinfo {author} {\bibnamefont{{Xue Zhike}}}, \bibinfo {author}
  {\bibnamefont{{Yan Xiaoli}}}, \bibinfo {author} {\bibnamefont{{Cheng Xin}}},
  \bibinfo {author} {\bibnamefont{{Yang Liheng}}}, \bibinfo {author}
  {\bibnamefont{{Su Yingna}}}, \bibinfo {author} {\bibnamefont{{Kliem
  Bernhard}}}, \bibinfo {author} {\bibnamefont{{Zhang Jun}}}, \bibinfo {author}
  {\bibnamefont{{Liu Zhong}}}, \bibinfo {author} {\bibnamefont{{Bi Yi}}},
  \bibinfo {author} {\bibnamefont{{Xiang Yongyuan}}}, \bibinfo {author}
  {\bibnamefont{{Yang Kai}}},\ and\ \bibinfo {author} {\bibnamefont{{Zhao
  Li}}},\ }%
  \bibfield{title}{%
  \enquote{\bibinfo {title} {{Observing the release of twist by magnetic
  reconnection in a solar filament eruption}},}\ }%
  \bibfield{journal}{%
  \bibinfo {journal} {Nat Commun}\ }%
  \textbf{\bibinfo {volume} {7}} (\bibinfo {month} {jun}\ \bibinfo {year}
  {2016}),\ \doi{\bibinfo {doi} {http://dx.doi.org/10.1038/ncomms11837
  10.1038/ncomms11837}}%
  \bibAnnoteFile{NoStop}{Zhike2016}%
\bibitem{Fazle&KarthikPof2011}%
  \BibitemOpen
  \bibfield{author}{%
  \bibinfo {author} {\bibfnamefont{Fazle}\ \bibnamefont{Hussain}}\ and\
  \bibinfo {author} {\bibfnamefont{Karthik}\ \bibnamefont{Duraisamy}},\ }%
  \bibfield{title}{%
  \enquote{\bibinfo {title} {Mechanics of viscous vortex reconnection},}\ }%
  \bibfield{journal}{%
  \bibinfo {journal} {Physics of Fluids}\ }%
  \textbf{\bibinfo {volume} {23}},\ \bibinfo {eid} {021701} (\bibinfo {year}
  {2011}),\ \doi{\bibinfo {doi} {http://dx.doi.org/10.1063/1.3532039}},\
  \url{http://scitation.aip.org/content/aip/journal/pof2/23/2/10.1063/1.353203%
9}%
  \bibAnnoteFile{NoStop}{Fazle&KarthikPof2011}%
\bibitem{Fonda25032014}%
  \BibitemOpen
  \bibfield{author}{%
  \bibinfo {author} {\bibfnamefont{Enrico}\ \bibnamefont{Fonda}}, \bibinfo
  {author} {\bibfnamefont{David~P.}\ \bibnamefont{Meichle}}, \bibinfo {author}
  {\bibfnamefont{Nicholas~T.}\ \bibnamefont{Ouellette}}, \bibinfo {author}
  {\bibfnamefont{Sahand}\ \bibnamefont{Hormoz}},\ and\ \bibinfo {author}
  {\bibfnamefont{Daniel~P.}\ \bibnamefont{Lathrop}},\ }%
  \bibfield{title}{%
  \enquote{\bibinfo {title} {Direct observation of kelvin waves excited by
  quantized vortex reconnection},}\ }%
  \bibfield{journal}{%
  \Doi{10.1073/pnas.1312536110}{\bibinfo {journal} {Proceedings of the National
  Academy of Sciences}}\ }%
  \textbf{\bibinfo {volume} {111}},\ \bibinfo {pages} {4707--4710} (\bibinfo
  {year} {2014}),\
  \url{http://www.pnas.org/content/111/Supplement_1/4707.abstract}%
  \bibAnnoteFile{NoStop}{Fonda25032014}%
\bibitem{constantin1996geometric}%
  \BibitemOpen
  \bibfield{author}{%
  \bibinfo {author} {\bibfnamefont{Peter}\ \bibnamefont{Constantin}}, \bibinfo
  {author} {\bibfnamefont{Charles}\ \bibnamefont{Fefferman}},\ and\ \bibinfo
  {author} {\bibfnamefont{Andrew~J}\ \bibnamefont{Majda}},\ }%
  \bibfield{title}{%
  \enquote{\bibinfo {title} {Geometric constraints on potentially singular
  solutions for the 3-d euler equations},}\ }%
  \bibfield{journal}{%
  \bibinfo {journal} {Communications in Partial Differential Equations}\ }%
  \textbf{\bibinfo {volume} {21}} (\bibinfo {year} {1996})%
  \bibAnnoteFile{NoStop}{constantin1996geometric}%
\bibitem{Moffatt2000}%
  \BibitemOpen
  \bibfield{author}{%
  \bibinfo {author} {\bibfnamefont{H.~K.}\ \bibnamefont{Moffatt}},\ }%
  \bibfield{title}{%
  \enquote{\bibinfo {title} {The interaction of skewed vortex pairs: a model
  for blow-up of the {Navier}--{Stokes} equations},}\ }%
  \bibfield{journal}{%
  \Doi{10.1017/S002211209900782X}{\bibinfo {journal} {Journal of Fluid
  Mechanics}}\ }%
  \textbf{\bibinfo {volume} {409}},\ \bibinfo {pages} {51--68} (\bibinfo
  {month} {04}\ \bibinfo {year} {2000}),\
  \url{https://www.cambridge.org/core/article/interaction-of-skewed-vortex-pai%
rs-a-model-for-blow-up-of-the-navier-stokes-equations/30C1C8B6C426DE7ECE26049E%
E188C105}%
  \bibAnnoteFile{NoStop}{Moffatt2000}%
\bibitem{Feynman195517}%
  \BibitemOpen
  \bibfield{author}{%
  \bibinfo {author} {\bibfnamefont{R.P.}\ \bibnamefont{Feynman}},\ }%
  \enquote{\bibinfo {title} {Chapter \{II\} application of quantum mechanics to
  liquid helium},}\ \ (\bibinfo {publisher} {Elsevier},\ \bibinfo {year}
  {1955})\ pp.\ \bibinfo {pages} {17 -- 53},\
  \url{http://www.sciencedirect.com/science/article/pii/S0079641708600773}%
  \bibAnnoteFile{NoStop}{Feynman195517}%
\bibitem{Schwarz_VF}%
  \BibitemOpen
  \bibfield{author}{%
  \bibinfo {author} {\bibfnamefont{K.~W.}\ \bibnamefont{Schwarz}},\ }%
  \bibfield{title}{%
  \enquote{\bibinfo {title} {Three-dimensional vortex dynamics in superfluid
  $^{4}\mathrm{He}$: Homogeneous superfluid turbulence},}\ }%
  \bibfield{journal}{%
  \Doi{10.1103/PhysRevB.38.2398}{\bibinfo {journal} {Phys. Rev. B}}\ }%
  \textbf{\bibinfo {volume} {38}},\ \bibinfo {pages} {2398--2417} (\bibinfo
  {month} {Aug}\ \bibinfo {year} {1988}),\
  \url{http://link.aps.org/doi/10.1103/PhysRevB.38.2398}%
  \bibAnnoteFile{NoStop}{Schwarz_VF}%
\bibitem{BewleyReconnectionPNAS}%
  \BibitemOpen
  \bibfield{author}{%
  \bibinfo {author} {\bibfnamefont{Gregory~P.}\ \bibnamefont{Bewley}}, \bibinfo
  {author} {\bibfnamefont{Matthew~S.}\ \bibnamefont{Paoletti}}, \bibinfo
  {author} {\bibfnamefont{Katepalli~R.}\ \bibnamefont{Sreenivasan}},\ and\
  \bibinfo {author} {\bibfnamefont{Daniel~P.}\ \bibnamefont{Lathrop}},\ }%
  \bibfield{title}{%
  \enquote{\bibinfo {title} {Characterization of reconnecting vortices in
  superfluid helium},}\ }%
  \bibfield{journal}{%
  \Doi{10.1073/pnas.0806002105}{\bibinfo {journal} {Proceedings of the National
  Academy of Sciences}}\ }%
  \textbf{\bibinfo {volume} {105}},\ \bibinfo {pages} {13707--13710} (\bibinfo
  {year} {2008}),\
  \Eprint{http://arxiv.org/abs/http://www.pnas.org/content/105/37/13707.full.p%
df}{http://www.pnas.org/content/105/37/13707.full.pdf},\
  \url{http://www.pnas.org/content/105/37/13707.abstract}%
  \bibAnnoteFile{NoStop}{BewleyReconnectionPNAS}%
\bibitem{PhysRevLett.86.2926}%
  \BibitemOpen
  \bibfield{author}{%
  \bibinfo {author} {\bibfnamefont{B.~P.}\ \bibnamefont{Anderson}}, \bibinfo
  {author} {\bibfnamefont{P.~C.}\ \bibnamefont{Haljan}}, \bibinfo {author}
  {\bibfnamefont{C.~A.}\ \bibnamefont{Regal}}, \bibinfo {author}
  {\bibfnamefont{D.~L.}\ \bibnamefont{Feder}}, \bibinfo {author}
  {\bibfnamefont{L.~A.}\ \bibnamefont{Collins}}, \bibinfo {author}
  {\bibfnamefont{C.~W.}\ \bibnamefont{Clark}},\ and\ \bibinfo {author}
  {\bibfnamefont{E.~A.}\ \bibnamefont{Cornell}},\ }%
  \bibfield{title}{%
  \enquote{\bibinfo {title} {Watching dark solitons decay into vortex rings in
  a bose-einstein condensate},}\ }%
  \bibfield{journal}{%
  \Doi{10.1103/PhysRevLett.86.2926}{\bibinfo {journal} {Phys. Rev. Lett.}}\ }%
  \textbf{\bibinfo {volume} {86}},\ \bibinfo {pages} {2926--2929} (\bibinfo
  {month} {Apr}\ \bibinfo {year} {2001}),\
  \url{http://link.aps.org/doi/10.1103/PhysRevLett.86.2926}%
  \bibAnnoteFile{NoStop}{PhysRevLett.86.2926}%
\bibitem{PhysRevLett.115.170402}%
  \BibitemOpen
  \bibfield{author}{%
  \bibinfo {author} {\bibfnamefont{S.}~\bibnamefont{Serafini}}, \bibinfo
  {author} {\bibfnamefont{M.}~\bibnamefont{Barbiero}}, \bibinfo {author}
  {\bibfnamefont{M.}~\bibnamefont{Debortoli}}, \bibinfo {author}
  {\bibfnamefont{S.}~\bibnamefont{Donadello}}, \bibinfo {author}
  {\bibfnamefont{F.}~\bibnamefont{Larcher}}, \bibinfo {author}
  {\bibfnamefont{F.}~\bibnamefont{Dalfovo}}, \bibinfo {author}
  {\bibfnamefont{G.}~\bibnamefont{Lamporesi}},\ and\ \bibinfo {author}
  {\bibfnamefont{G.}~\bibnamefont{Ferrari}},\ }%
  \bibfield{title}{%
  \enquote{\bibinfo {title} {Dynamics and interaction of vortex lines in an
  elongated bose-einstein condensate},}\ }%
  \bibfield{journal}{%
  \Doi{10.1103/PhysRevLett.115.170402}{\bibinfo {journal} {Phys. Rev. Lett.}}\
  }%
  \textbf{\bibinfo {volume} {115}},\ \bibinfo {pages} {170402} (\bibinfo
  {month} {Oct}\ \bibinfo {year} {2015}),\
  \url{http://link.aps.org/doi/10.1103/PhysRevLett.115.170402}%
  \bibAnnoteFile{NoStop}{PhysRevLett.115.170402}%
\bibitem{Baggaley&Sherwin&Barenghi&SergeePRB2012}%
  \BibitemOpen
  \bibfield{author}{%
  \bibinfo {author} {\bibfnamefont{A.~W.}\ \bibnamefont{Baggaley}}, \bibinfo
  {author} {\bibfnamefont{L.~K.}\ \bibnamefont{Sherwin}}, \bibinfo {author}
  {\bibfnamefont{C.~F.}\ \bibnamefont{Barenghi}},\ and\ \bibinfo {author}
  {\bibfnamefont{Y.~A.}\ \bibnamefont{Sergeev}},\ }%
  \bibfield{title}{%
  \enquote{\bibinfo {title} {Thermally and mechanically driven quantum
  turbulence in helium {II}},}\ }%
  \bibfield{journal}{%
  \Doi{10.1103/PhysRevB.86.104501}{\bibinfo {journal} {Phys. Rev. B}}\ }%
  \textbf{\bibinfo {volume} {86}},\ \bibinfo {pages} {104501} (\bibinfo {month}
  {Sep}\ \bibinfo {year} {2012}),\
  \url{http://link.aps.org/doi/10.1103/PhysRevB.86.104501}%
  \bibAnnoteFile{NoStop}{Baggaley&Sherwin&Barenghi&SergeePRB2012}%
\bibitem{nazarenko2003analytical}%
  \BibitemOpen
  \bibfield{author}{%
  \bibinfo {author} {\bibfnamefont{Sergey}\ \bibnamefont{Nazarenko}}\ and\
  \bibinfo {author} {\bibfnamefont{Robert}\ \bibnamefont{West}},\ }%
  \bibfield{title}{%
  \enquote{\bibinfo {title} {Analytical solution for nonlinear
  {Schr{\"o}dinger} vortex reconnection},}\ }%
  \bibfield{journal}{%
  \bibinfo {journal} {Journal of low temperature physics}\ }%
  \textbf{\bibinfo {volume} {132}},\ \bibinfo {pages} {1--10} (\bibinfo {year}
  {2003})%
  \bibAnnoteFile{NoStop}{nazarenko2003analytical}%
\bibitem{Zuccher&Caliari&Baggaley&BarenghiPof2012}%
  \BibitemOpen
  \bibfield{author}{%
  \bibinfo {author} {\bibfnamefont{S.}~\bibnamefont{Zuccher}}, \bibinfo
  {author} {\bibfnamefont{M.}~\bibnamefont{Caliari}}, \bibinfo {author}
  {\bibfnamefont{A.~W.}\ \bibnamefont{Baggaley}},\ and\ \bibinfo {author}
  {\bibfnamefont{C.~F.}\ \bibnamefont{Barenghi}},\ }%
  \bibfield{title}{%
  \enquote{\bibinfo {title} {Quantum vortex reconnections},}\ }%
  \bibfield{journal}{%
  \bibinfo {journal} {Physics of Fluids}\ }%
  \textbf{\bibinfo {volume} {24}},\ \bibinfo {eid} {125108} (\bibinfo {year}
  {2012}),\ \doi{\bibinfo {doi} {http://dx.doi.org/10.1063/1.4772198}},\
  \url{http://scitation.aip.org/content/aip/journal/pof2/24/12/10.1063/1.47721%
98}%
  \bibAnnoteFile{NoStop}{Zuccher&Caliari&Baggaley&BarenghiPof2012}%
\bibitem{AllenFiniteTempRecoPRA2014}%
  \BibitemOpen
  \bibfield{author}{%
  \bibinfo {author} {\bibfnamefont{A.~J.}\ \bibnamefont{Allen}}, \bibinfo
  {author} {\bibfnamefont{S.}~\bibnamefont{Zuccher}}, \bibinfo {author}
  {\bibfnamefont{M.}~\bibnamefont{Caliari}}, \bibinfo {author}
  {\bibfnamefont{N.~P.}\ \bibnamefont{Proukakis}}, \bibinfo {author}
  {\bibfnamefont{N.~G.}\ \bibnamefont{Parker}},\ and\ \bibinfo {author}
  {\bibfnamefont{C.~F.}\ \bibnamefont{Barenghi}},\ }%
  \bibfield{title}{%
  \enquote{\bibinfo {title} {Vortex reconnections in atomic condensates at
  finite temperature},}\ }%
  \bibfield{journal}{%
  \Doi{10.1103/PhysRevA.90.013601}{\bibinfo {journal} {Phys. Rev. A}}\ }%
  \textbf{\bibinfo {volume} {90}},\ \bibinfo {pages} {013601} (\bibinfo {month}
  {Jul}\ \bibinfo {year} {2014}),\
  \url{http://link.aps.org/doi/10.1103/PhysRevA.90.013601}%
  \bibAnnoteFile{NoStop}{AllenFiniteTempRecoPRA2014}%
\bibitem{2014arXiv1410.1259R}%
  \BibitemOpen
  \bibfield{author}{%
  \bibinfo {author} {\bibfnamefont{C.}~\bibnamefont{Rorai}}, \bibinfo {author}
  {\bibfnamefont{J.}~\bibnamefont{Skipper}}, \bibinfo {author}
  {\bibfnamefont{R.~M.}\ \bibnamefont{Kerr}},\ and\ \bibinfo {author}
  {\bibfnamefont{K.~R.}\ \bibnamefont{Sreenivasan}},\ }%
  \bibfield{title}{%
  \enquote{\bibinfo {title} {Approach and separation of quantised vortices with
  balanced cores},}\ }%
  \bibfield{journal}{%
  \Doi{10.1017/jfm.2016.638}{\bibinfo {journal} {Journal of Fluid Mechanics}}\
  }%
  \textbf{\bibinfo {volume} {808}},\ \bibinfo {pages} {641?667} (\bibinfo
  {month} {Dec}\ \bibinfo {year} {2016}),\
  \url{https://www.cambridge.org/core/article/approach-and-separation-of-quant%
ised-vortices-with-balanced-cores/707BC6C045FC2360D86DD0CFD4D48081}%
  \bibAnnoteFile{NoStop}{2014arXiv1410.1259R}%
\bibitem{Waele&AartsPRL1994}%
  \BibitemOpen
  \bibfield{author}{%
  \bibinfo {author} {\bibfnamefont{A.~T. A.~M.}\ \bibnamefont{de~Waele}}\ and\
  \bibinfo {author} {\bibfnamefont{R.~G. K.~M.}\ \bibnamefont{Aarts}},\ }%
  \bibfield{title}{%
  \enquote{\bibinfo {title} {Route to vortex reconnection},}\ }%
  \bibfield{journal}{%
  \Doi{10.1103/PhysRevLett.72.482}{\bibinfo {journal} {Phys. Rev. Lett.}}\ }%
  \textbf{\bibinfo {volume} {72}},\ \bibinfo {pages} {482--485} (\bibinfo
  {month} {Jan}\ \bibinfo {year} {1994}),\
  \url{http://link.aps.org/doi/10.1103/PhysRevLett.72.482}%
  \bibAnnoteFile{NoStop}{Waele&AartsPRL1994}%
\bibitem{tebbs2011approach}%
  \BibitemOpen
  \bibfield{author}{%
  \bibinfo {author} {\bibfnamefont{Richard}\ \bibnamefont{Tebbs}}, \bibinfo
  {author} {\bibfnamefont{Anthony~J}\ \bibnamefont{Youd}},\ and\ \bibinfo
  {author} {\bibfnamefont{Carlo~F}\ \bibnamefont{Barenghi}},\ }%
  \bibfield{title}{%
  \enquote{\bibinfo {title} {The approach to vortex reconnection},}\ }%
  \bibfield{journal}{%
  \bibinfo {journal} {Journal of Low Temperature Physics}\ }%
  \textbf{\bibinfo {volume} {162}},\ \bibinfo {pages} {314--321} (\bibinfo
  {year} {2011})%
  \bibAnnoteFile{NoStop}{tebbs2011approach}%
\bibitem{PhysRevLett.86.3080}%
  \BibitemOpen
  \bibfield{author}{%
  \bibinfo {author} {\bibfnamefont{D.}~\bibnamefont{Kivotides}}, \bibinfo
  {author} {\bibfnamefont{J.~C.}\ \bibnamefont{Vassilicos}}, \bibinfo {author}
  {\bibfnamefont{D.~C.}\ \bibnamefont{Samuels}},\ and\ \bibinfo {author}
  {\bibfnamefont{C.~F.}\ \bibnamefont{Barenghi}},\ }%
  \bibfield{title}{%
  \enquote{\bibinfo {title} {Kelvin waves cascade in superfluid turbulence},}\
  }%
  \bibfield{journal}{%
  \Doi{10.1103/PhysRevLett.86.3080}{\bibinfo {journal} {Phys. Rev. Lett.}}\ }%
  \textbf{\bibinfo {volume} {86}},\ \bibinfo {pages} {3080--3083} (\bibinfo
  {month} {Apr}\ \bibinfo {year} {2001}),\
  \url{http://link.aps.org/doi/10.1103/PhysRevLett.86.3080}%
  \bibAnnoteFile{NoStop}{PhysRevLett.86.3080}%
\bibitem{Nazarenko2007}%
  \BibitemOpen
  \bibfield{author}{%
  \bibinfo {author} {\bibfnamefont{S.}~\bibnamefont{Nazarenko}},\ }%
  \bibfield{title}{%
  \enquote{\bibinfo {title} {Kelvin wave turbulence generated by vortex
  reconnections},}\ }%
  \bibfield{journal}{%
  \Doi{10.1134/S0021364006230032}{\bibinfo {journal} {JETP Letters}}\ }%
  \textbf{\bibinfo {volume} {84}},\ \bibinfo {pages} {585--587} (\bibinfo
  {year} {2007}),\ ISSN \bibinfo {issn} {1090-6487},\
  \url{http://dx.doi.org/10.1134/S0021364006230032}%
  \bibAnnoteFile{NoStop}{Nazarenko2007}%
\bibitem{PhysRevB.92.184508}%
  \BibitemOpen
  \bibfield{author}{%
  \bibinfo {author} {\bibfnamefont{R.}~\bibnamefont{H\"anninen}},\ }%
  \bibfield{title}{%
  \enquote{\bibinfo {title} {Kelvin waves from vortex reconnection in
  superfluid helium at low temperatures},}\ }%
  \bibfield{journal}{%
  \Doi{10.1103/PhysRevB.92.184508}{\bibinfo {journal} {Phys. Rev. B}}\ }%
  \textbf{\bibinfo {volume} {92}},\ \bibinfo {pages} {184508} (\bibinfo {month}
  {Nov}\ \bibinfo {year} {2015}),\
  \url{http://link.aps.org/doi/10.1103/PhysRevB.92.184508}%
  \bibAnnoteFile{NoStop}{PhysRevB.92.184508}%
\bibitem{Scheeler28102014PNASDavide}%
  \BibitemOpen
  \bibfield{author}{%
  \bibinfo {author} {\bibfnamefont{Martin~W.}\ \bibnamefont{Scheeler}},
  \bibinfo {author} {\bibfnamefont{Dustin}\ \bibnamefont{Kleckner}}, \bibinfo
  {author} {\bibfnamefont{Davide}\ \bibnamefont{Proment}}, \bibinfo {author}
  {\bibfnamefont{Gordon~L.}\ \bibnamefont{Kindlmann}},\ and\ \bibinfo {author}
  {\bibfnamefont{William T.~M.}\ \bibnamefont{Irvine}},\ }%
  \bibfield{title}{%
  \enquote{\bibinfo {title} {Helicity conservation by flow across scales in
  reconnecting vortex links and knots},}\ }%
  \bibfield{journal}{%
  \Doi{10.1073/pnas.1407232111}{\bibinfo {journal} {Proceedings of the National
  Academy of Sciences}}\ }%
  \textbf{\bibinfo {volume} {111}},\ \bibinfo {pages} {15350--15355} (\bibinfo
  {year} {2014}),\
  \Eprint{http://arxiv.org/abs/http://www.pnas.org/content/111/43/15350.full.p%
df}{http://www.pnas.org/content/111/43/15350.full.pdf},\
  \url{http://www.pnas.org/content/111/43/15350.abstract}%
  \bibAnnoteFile{NoStop}{Scheeler28102014PNASDavide}%
\bibitem{kimura2014}%
  \BibitemOpen
  \bibfield{author}{%
  \bibinfo {author} {\bibfnamefont{Y.}~\bibnamefont{Kimura}}\ and\ \bibinfo
  {author} {\bibfnamefont{H.Â~K.}\ \bibnamefont{Moffatt}},\ }%
  \bibfield{title}{%
  \enquote{\bibinfo {title} {Reconnection of skewed vortices},}\ }%
  \bibfield{journal}{%
  \Doi{10.1017/jfm.2014.233}{\bibinfo {journal} {Journal of Fluid Mechanics}}\
  }%
  \textbf{\bibinfo {volume} {751}},\ \bibinfo {pages} {329--345} (\bibinfo
  {month} {007}\ \bibinfo {year} {2014}),\
  \url{https://www.cambridge.org/core/article/reconnection-of-skewed-vortices/%
27FA228C40A837AA9744B9716FDB3A07}%
  \bibAnnoteFile{NoStop}{kimura2014}%
\bibitem{laing2015conservation}%
  \BibitemOpen
  \bibfield{author}{%
  \bibinfo {author} {\bibfnamefont{Christian~E}\ \bibnamefont{Laing}}, \bibinfo
  {author} {\bibfnamefont{Renzo~L}\ \bibnamefont{Ricca}},\ and\ \bibinfo
  {author} {\bibfnamefont{L~Sumners}\ \bibnamefont{De~Witt}},\ }%
  \bibfield{title}{%
  \enquote{\bibinfo {title} {Conservation of writhe helicity under
  anti-parallel reconnection},}\ }%
  \bibfield{journal}{%
  \bibinfo {journal} {Scientific reports}\ }%
  \textbf{\bibinfo {volume} {5}} (\bibinfo {year} {2015})%
  \bibAnnoteFile{NoStop}{laing2015conservation}%
\bibitem{Zuccher&RiccaPRE2015}%
  \BibitemOpen
  \bibfield{author}{%
  \bibinfo {author} {\bibfnamefont{Simone}\ \bibnamefont{Zuccher}}\ and\
  \bibinfo {author} {\bibfnamefont{Renzo~L.}\ \bibnamefont{Ricca}},\ }%
  \bibfield{title}{%
  \enquote{\bibinfo {title} {Helicity conservation under quantum reconnection
  of vortex rings},}\ }%
  \bibfield{journal}{%
  \Doi{10.1103/PhysRevE.92.061001}{\bibinfo {journal} {Phys. Rev. E}}\ }%
  \textbf{\bibinfo {volume} {92}},\ \bibinfo {pages} {061001} (\bibinfo {month}
  {Dec}\ \bibinfo {year} {2015}),\
  \url{http://link.aps.org/doi/10.1103/PhysRevE.92.061001}%
  \bibAnnoteFile{NoStop}{Zuccher&RiccaPRE2015}%
\bibitem{diLeoniHelicity}%
  \BibitemOpen
  \bibfield{author}{%
  \bibinfo {author} {\bibfnamefont{P.}~\bibnamefont{Clark~di Leoni}}, \bibinfo
  {author} {\bibfnamefont{P.~D.}\ \bibnamefont{Mininni}},\ and\ \bibinfo
  {author} {\bibfnamefont{M.~E.}\ \bibnamefont{Brachet}},\ }%
  \bibfield{title}{%
  \enquote{\bibinfo {title} {Helicity, topology, and kelvin waves in
  reconnecting quantum knots},}\ }%
  \bibfield{journal}{%
  \Doi{10.1103/PhysRevA.94.043605}{\bibinfo {journal} {Phys. Rev. A}}\ }%
  \textbf{\bibinfo {volume} {94}},\ \bibinfo {pages} {043605} (\bibinfo {month}
  {Oct}\ \bibinfo {year} {2016}),\
  \url{http://link.aps.org/doi/10.1103/PhysRevA.94.043605}%
  \bibAnnoteFile{NoStop}{diLeoniHelicity}%
\bibitem{VilloisTrackingAlgo}%
  \BibitemOpen
  \bibfield{author}{%
  \bibinfo {author} {\bibfnamefont{Alberto}\ \bibnamefont{Villois}}, \bibinfo
  {author} {\bibfnamefont{Giorgio}\ \bibnamefont{Krstulovic}}, \bibinfo
  {author} {\bibfnamefont{Davide}\ \bibnamefont{Proment}},\ and\ \bibinfo
  {author} {\bibfnamefont{Hayder}\ \bibnamefont{Salman}},\ }%
  \bibfield{title}{%
  \enquote{\bibinfo {title} {A vortex filament tracking method for the
  {Gross}--{Pitaevskii} model of a superfluid},}\ }%
  \bibfield{journal}{%
  \bibinfo {journal} {Journal of Physics A: Mathematical and Theoretical}\ }%
  \textbf{\bibinfo {volume} {49}},\ \bibinfo {pages} {415502} (\bibinfo {year}
  {2016}),\ \url{http://stacks.iop.org/1751-8121/49/i=41/a=415502}%
  \bibAnnoteFile{NoStop}{VilloisTrackingAlgo}%
\bibitem{Berloff2001}%
  \BibitemOpen
  \bibfield{author}{%
  \bibinfo {author} {\bibfnamefont{Natalia~G}\ \bibnamefont{Berloff}}\ and\
  \bibinfo {author} {\bibfnamefont{Paul~H}\ \bibnamefont{Roberts}},\ }%
  \bibfield{title}{%
  \enquote{\bibinfo {title} {Motion in a bose condensate: Ix. crow instability
  of antiparallel vortex pairs},}\ }%
  \bibfield{journal}{%
  \bibinfo {journal} {Journal of Physics A: Mathematical and General}\ }%
  \textbf{\bibinfo {volume} {34}},\ \bibinfo {pages} {10057} (\bibinfo {year}
  {2001}),\ \url{http://stacks.iop.org/0305-4470/34/i=47/a=311}%
  \bibAnnoteFile{NoStop}{Berloff2001}%
\bibitem{Proment2012}%
  \BibitemOpen
  \bibfield{author}{%
  \bibinfo {author} {\bibfnamefont{Davide}\ \bibnamefont{Proment}}, \bibinfo
  {author} {\bibfnamefont{Miguel}\ \bibnamefont{Onorato}},\ and\ \bibinfo
  {author} {\bibfnamefont{Carlo~F.}\ \bibnamefont{Barenghi}},\ }%
  \bibfield{title}{%
  \enquote{\bibinfo {title} {Vortex knots in a bose-einstein condensate},}\ }%
  \bibfield{journal}{%
  \Doi{10.1103/PhysRevE.85.036306}{\bibinfo {journal} {Phys. Rev. E}}\ }%
  \textbf{\bibinfo {volume} {85}},\ \bibinfo {pages} {036306} (\bibinfo {month}
  {Mar}\ \bibinfo {year} {2012}),\
  \url{http://link.aps.org/doi/10.1103/PhysRevE.85.036306}%
  \bibAnnoteFile{NoStop}{Proment2012}%
\bibitem{nore1997decaying}%
  \BibitemOpen
  \bibfield{author}{%
  \bibinfo {author} {\bibfnamefont{Caroline}\ \bibnamefont{Nore}}, \bibinfo
  {author} {\bibfnamefont{Malek}\ \bibnamefont{Abid}},\ and\ \bibinfo {author}
  {\bibfnamefont{ME}~\bibnamefont{Brachet}},\ }%
  \bibfield{title}{%
  \enquote{\bibinfo {title} {Decaying {Kolmogorov} turbulence in a model of
  superflow},}\ }%
  \bibfield{journal}{%
  \bibinfo {journal} {Physics of Fluids (1994-present)}\ }%
  \textbf{\bibinfo {volume} {9}},\ \bibinfo {pages} {2644--2669} (\bibinfo
  {year} {1997})%
  \bibAnnoteFile{NoStop}{nore1997decaying}%
\bibitem{VilloisTangleLetter}%
  \BibitemOpen
  \bibfield{author}{%
  \bibinfo {author} {\bibfnamefont{Alberto}\ \bibnamefont{Villois}}, \bibinfo
  {author} {\bibfnamefont{Davide}\ \bibnamefont{Proment}},\ and\ \bibinfo
  {author} {\bibfnamefont{Giorgio}\ \bibnamefont{Krstulovic}},\ }%
  \bibfield{title}{%
  \enquote{\bibinfo {title} {Evolution of a superfluid vortex filament tangle
  driven by the {Gross}--{Pitaevskii} equation},}\ }%
  \bibfield{journal}{%
  \Doi{10.1103/PhysRevE.93.061103}{\bibinfo {journal} {Phys. Rev. E}}\ }%
  \textbf{\bibinfo {volume} {93}},\ \bibinfo {pages} {061103} (\bibinfo {month}
  {Jun}\ \bibinfo {year} {2016}),\
  \url{http://link.aps.org/doi/10.1103/PhysRevE.93.061103}%
  \bibAnnoteFile{NoStop}{VilloisTangleLetter}%
\bibitem{PhysRevE.51.3207}%
  \BibitemOpen
  \bibfield{author}{%
  \bibinfo {author} {\bibfnamefont{Peter}\ \bibnamefont{Constantin}}, \bibinfo
  {author} {\bibfnamefont{Itamar}\ \bibnamefont{Procaccia}},\ and\ \bibinfo
  {author} {\bibfnamefont{Daniel}\ \bibnamefont{Segel}},\ }%
  \bibfield{title}{%
  \enquote{\bibinfo {title} {Creation and dynamics of vortex tubes in
  three-dimensional turbulence},}\ }%
  \bibfield{journal}{%
  \Doi{10.1103/PhysRevE.51.3207}{\bibinfo {journal} {Phys. Rev. E}}\ }%
  \textbf{\bibinfo {volume} {51}},\ \bibinfo {pages} {3207--3222} (\bibinfo
  {month} {Apr}\ \bibinfo {year} {1995}),\
  \url{http://link.aps.org/doi/10.1103/PhysRevE.51.3207}%
  \bibAnnoteFile{NoStop}{PhysRevE.51.3207}%
\bibitem{serafini2016vortex}%
  \BibitemOpen
  \bibfield{author}{%
  \bibinfo {author} {\bibfnamefont{Simone}\ \bibnamefont{Serafini}}, \bibinfo
  {author} {\bibfnamefont{Luca}\ \bibnamefont{Galantucci}}, \bibinfo {author}
  {\bibfnamefont{Elena}\ \bibnamefont{Iseni}}, \bibinfo {author}
  {\bibfnamefont{Tom}\ \bibnamefont{Bienaim{\'e}}}, \bibinfo {author}
  {\bibfnamefont{Russell~N}\ \bibnamefont{Bisset}}, \bibinfo {author}
  {\bibfnamefont{Carlo~F}\ \bibnamefont{Barenghi}}, \bibinfo {author}
  {\bibfnamefont{Franco}\ \bibnamefont{Dalfovo}}, \bibinfo {author}
  {\bibfnamefont{Giacomo}\ \bibnamefont{Lamporesi}},\ and\ \bibinfo {author}
  {\bibfnamefont{Gabriele}\ \bibnamefont{Ferrari}},\ }%
  \bibfield{title}{%
  \enquote{\bibinfo {title} {Vortex reconnections and rebounds in trapped
  atomic bose--einstein condensates},}\ }%
  \bibfield{journal}{%
  \bibinfo {journal} {arXiv preprint arXiv:1611.01691}}%
   (\bibinfo {year} {2016})%
  \bibAnnoteFile{NoStop}{serafini2016vortex}%
\end{thebibliography}%

\end{document}